\documentclass[aps,prd,amsmath,amssymb,preprintnumbers,groupedaddress,twocolumn,nofootinbib]{revtex4-1}
\usepackage{tikz}
\usepackage{slashed}
\usepackage{hyperref}
\usepackage{verbatim}
\usepackage[normalem]{ulem}

\newcommand{\gev}{\text{GeV}}
\newcommand{\mev}{\text{MeV}}
\newcommand{\tev}{\text{TeV}}

\newcommand{\trh}{T_\text{RH}}
\newcommand{\NDW}{N_\text{DW}}

\newcommand{\lsim}{\!\mathrel{\hbox{\rlap{\lower.55ex \hbox{$\sim$}} \kern-.34em \raise.4ex \hbox{$<$}}}}
\newcommand{\gsim}{\!\mathrel{\hbox{\rlap{\lower.55ex \hbox{$\sim$}} \kern-.34em \raise.4ex \hbox{$>$}}}}
\newcommand{\hc}{\text{ h.c.}}
\newcommand{\vev}[1]{ \left\langle \abs{#1} \right\rangle }
\newcommand{\vevnoabs}[1]{ \left\langle #1 \right\rangle }
\newcommand{\abs}[1]{ \left| {#1} \right| }
\def\be{\begin{equation}}
\def\ee{\end{equation}}
\newcommand{\Eref}[1]{Eq.~(\ref{#1})}
\newcommand{\Erefs}[2]{Eqs.~(\ref{#1}) and~(\ref{#2})}
\newcommand{\Fref}[1]{Fig.~\ref{#1}}
\newcommand{\Sref}[1]{Sec.~\ref{#1}}

\def\ie{{\it i.e.}}
\def\eg{{\it e.g.}}

\begin{document}
\setlength{\baselineskip}{0.22in}

\preprint{FERMILAB-PUB-16-005-T}
\preprint{MCTP-16-02}
\title{High-Scale Axions without Isocurvature from Inflationary Dynamics}
\author{John Kearney} \affiliation{Theoretical Physics Department, Fermi National Accelerator Laboratory\\ Batavia, IL 60510 USA} 
\author{Nicholas Orlofsky and Aaron Pierce} \affiliation{Michigan Center for Theoretical Physics (MCTP), Department of Physics, University of Michigan \\Ann Arbor, MI 48109 USA}
\date{\today}

\begin{abstract}
Observable primordial tensor modes in the cosmic microwave background (CMB) would point to a high scale of inflation $H_{I}$.  If the scale of Peccei-Quinn (PQ) breaking $f_a$ is greater than $\frac{H_{I}}{2\pi}$, CMB constraints on isocurvature na\"ively rule out QCD axion dark matter.  This assumes the potential of the axion is unmodified during inflation.  We revisit models where inflationary dynamics modify the axion potential and discuss how isocurvature bounds can be relaxed.  We find that models that rely solely on a larger PQ-breaking scale during inflation $f_I$ require either late-time dilution of the axion abundance or highly super-Planckian $f_I$ that somehow does not dominate the inflationary energy density.
Models that have enhanced explicit breaking of the PQ symmetry during inflation may allow $f_a$ close to the Planck scale.
Avoiding disruption of inflationary dynamics provides important limits on the parameter space.
\end{abstract}
\maketitle
\section{Introduction}

The axion \cite{Weinberg:1977ma, Wilczek:1977pj} is an elegant solution to the strong CP problem \cite{Peccei:1977hh}. 
It arises as a pseudo-Nambu-Goldstone boson (pNGB) of a spontaneously broken Peccei-Quinn (PQ) symmetry at a scale $f_a$.  If its potential is dominated by QCD instanton effects, the axion settles into an approximately CP-conserving minimum.
In addition, the axion is an attractive candidate for the dark matter (DM) in our universe \cite{Abbott:1982af,Preskill:1982cy,Dine:1982ah}.

Axion cosmology depends on whether the PQ symmetry is spontaneously broken before inflation ends.  
The Hubble parameter during inflation $H_I$ can be related to the tensor perturbation amplitude in the cosmic microwave background (CMB) \cite{Abbott:1984fp}.  Current bounds from BICEP2/Keck/Planck constrain the tensor-to-scalar ratio $r<.07$ \cite{Array:2015xqh,Ade:2015lrj}, while planned near-future detectors will probe the region $r \gsim 2 \times 10^{-3}$ \cite{Creminelli:2015oda}.
If such a primordial gravity wave detection is made by these experiments, it will indicate 
\begin{equation}
H_I = \left(\frac{\pi^2}{2}M_P^2 \Delta_R^2 r\right)^{1/2} \gsim 10^{13}~\gev,
\end{equation} 
where the measured scalar perturbation amplitude $\Delta_R^2=2.142 \times 10^{-9}$ \cite{Ade:2015xua} and the reduced Planck mass $M_P=2.435 \times 10^{18}~\gev$.

If the PQ symmetry breaks after inflation ($f_a < \frac{H_I}{2 \pi}$), topological defects will form.  These topological defects will lead to overclosure due to either domain wall stability if the domain wall number $\NDW>1$, or axion overproduction from cosmic strings and domain walls if $f_a \gsim 10^{11}~\gev$ \cite{Harari:1987ht,*Chang:1998bq,*Hiramatsu:2012gg,*Davis:1986xc,*Battye:1993jv,*Battye:1994au}.
Larger PQ-breaking scales and domain wall numbers other than one are allowed only if the PQ symmetry is broken before the end of inflation ($f_a> \frac{H_I}{2\pi}$) so that topological defects are inflated away.\footnote{Alternatively, topological defects may be destabilized via a very delicately chosen tilt to the  potential \cite{Kawasaki:2014sqa,Ringwald:2015dsf}.}
However, in this case the (massless) axion will obtain inflationary fluctuations $\sim \frac{H_I}{2 \pi}$ similar to the inflaton \cite{Lyth:1992tx,Lyth:1991ub,Lyth:1989pb,Linde:1991km}.  Unlike the inflaton fluctuations, the fluctuations of the axion would today appear as isocurvature and are bounded by observations of the CMB \cite{Ade:2015xua}.

Narrowing to the case that the PQ symmetry is broken before the end of inflation, once inflation ends the axial component of the PQ field remains at the value taken during inflation until the temperature-dependent QCD instanton axion mass $m_{\rm QCD}^2$ becomes comparable to the Hubble expansion rate (typically near the QCD phase transition).
At this point, the field will begin oscillating around the minimum of the low-energy potential.
These coherent oscillations correspond to an energy density stored in the axion field, meaning that the axion can behave as cold DM.  The QCD axion DM fraction generated in this manner is given by \cite{Kolb:1990vq, Visinelli:2014twa}
\be
\label{eq:axionfraction}
R_a= \frac{\Omega_a}{\Omega_\text{DM}} \simeq \left\{
\begin{array}{ll}
5.6 \times 10^7 \vevnoabs{\theta^2} (f_a/M_P)^{7/6}, & f_a<\hat{f}_a, \\
1.6 \times 10^8 \vevnoabs{\theta^2} (f_a/M_P)^{3/2}, & f_a>\hat{f}_a,
\end{array}
\right.
\ee
where $\vevnoabs{\theta^2} = \theta_i^2+\delta \theta^2$ is the axion misalignment angle including contributions from the initial misalignment $\theta_i$ and the primordial fluctuations $\delta \theta$ (neglecting anharmonic factors \cite{Turner:1985si}).
The scale $\hat{f}_a \simeq .991 \times 10^{17}~\gev \simeq \frac{M_P}{2.5}$ corresponds to the transition between oscillations beginning before or after the QCD phase transition (see \cite{Kitano:2015fla,diCortona:2015ldu} for corrections to this approximation).

Using this expression for the axion DM fraction and assuming a detection of primordial tensor modes---indicating a large $H_I$ and thus large isocurvature fluctuations $\delta \theta$---current constraints on isocurvature would rule out the simplest axion models for $f_{a} > \frac{H_I}{2\pi}$, see, \eg, \cite{Fox:2004kb,Visinelli:2014twa}.  For high-scale axion models to be viable in the presence of such a detection, the axion fluctuations must be suppressed.

It is important to understand the robustness of isocurvature constraints for two reasons.
First, string theories generically predict a PQ-breaking scale around the grand unified theory (GUT) scale, $f_a \sim 10^{16}~\gev$ \cite{Choi:1985bz, Banks:1996ea}. Second, several experiments have recently been proposed to look for large-$f_a$ axion DM \cite{Budker:2013hfa, Sikivie:2013laa, Sikivie:2014lha, Popov:2014mba}.  If primordial gravity waves are observed, it will be important to know whether such theories are permissible and thus which regions of parameter space these experiments should target.
More optimistically, if signals are seen both in tensor modes and in a hunt for large-$f_{a}$ axions, we should know what types of new physics would be required to reconcile the two signals.

We examine two classes of solutions that have been discussed to circumvent isocurvature constraints and resurrect models with large $f_a$.  Both involve an inflationary shift of the PQ sector away from its zero temperature minimum and/or potential.  In the first, the PQ scale is large during inflation $f_{I} \gg f_{a}$   \cite{Linde:1990yj,*Linde:1991km,Higaki:2014ooa,Chun:2014xva,Fairbairn:2014zta}.  In the second, the  axion has a large mass during inflation related to an explicit PQ symmetry breaking \cite{Dine:2004cq,Higaki:2014ooa}.\footnote{Other mechanisms to suppress isocurvature have also been proposed \cite{Dvali:1995ce,Folkerts:2013tua,Jeong:2013xta,Choi:2014uaa,Kawasaki:2014una,Choi:2015zra,Nakayama:2015pba,Takahashi:2015waa,Kawasaki:2015lpf,Nomura:2015xil}, though notably many of these do not work for GUT-scale axion DM with high-scale inflation.}
We explore the measures necessary to suppress isocurvature and the constraints that must be taken into account as these measures are implemented.

In Section \ref{sec:wfsuppression}, we discuss the viability of exclusively suppressing isocurvature via an inflationary PQ-breaking scale that is larger than the present scale.  While this has been previously considered as a viable mechanism to reduce isocurvature, we show that it often faces insurmountable constraints due to the large hierarchy required between the two scales.
In Section \ref{sec:axionmass}, we explore the possibility that isocurvature can be suppressed with an enhanced explicit breaking of the PQ symmetry, demonstrating the model-building and experimental constraints that ultimately limit the range that $f_a$ may take in such models.
Numerous previous works have observed that explicit PQ-breaking modifies the axion potential, and so in general could be a suitable strategy for suppressing isocurvature.
Here, however, we focus on developing concrete models, which allows us to study the various interrelated field dynamics and constraints, and we are thus able to more fully address the strengths and shortcomings of this approach.

As we are interested in modifying the behavior of the axion during inflation, our primary focus here will be on how the solutions implemented affect the dynamics of the axion and inflaton fields.
However, the solutions under consideration in the above sections can also affect the behavior of the radial component of the PQ-breaking field.  This leads to model-dependent, but potentially severe, constraints related to ensuring a consistent cosmological history.  We review these issues in Section \ref{sec:additionalconstraints}.
Many of these constraints could be evaded if the PQ field relaxed adiabatically to its minimum today without oscillation. However, in Section \ref{sec:adiabatic}, we discuss how models that are designed to permit such adiabatic relaxation will generally only partially mitigate constraints.
Moreover, we highlight some previously unappreciated obstacles to constructing such models, which make it unclear whether adiabatic relaxation can in fact be implemented.
We conclude in Section \ref{sec:conclusion}.

 
\section{Suppressing Isocurvature via Wave Function Renormalization}
\label{sec:wfsuppression}

The complex PQ field $S$ can be written in terms of its radial and axial components, $\sigma$ and $a$, and the PQ breaking scale, $f_{\rm PQ}$,
\be 
\label{eqn:Snonlin}
S=\frac{1}{\sqrt{2}}\left(\sigma+f\right) \exp\left[\frac{ia}{f}\right],
\ee
where $f = \frac{f_{\rm PQ}}{\NDW}$, with $f = f_a$ and $f = f_I$ denoting the effective breaking scale today and during inflation, respectively.
During inflation, fluctuations $\sim \frac{H_I}{2\pi}$ are induced in the canonically normalized axion field,
which correspond to fluctuations in the field today given by \cite{Linde:1990yj,*Linde:1991km},
\be
\delta a=\frac{H_I f_a}{2 \pi f_I}.
\ee
Thus, $f_I > f_a$ will reduce axion fluctuations.
However, due to the stringent bounds on isocurvature, $\alpha_\text{iso}<.0019$ at the 95\% confidence level \cite{Ade:2015xua}, $f_I \gg f_a$ is necessary to achieve the required suppression.
The isocurvature parameter is
\be
\frac{\alpha_\text{iso}}{1-\alpha_\text{iso}} \simeq \frac{1}{\Delta_R^2} \left(4 R_a^2 \frac{(H_I/(2 \pi))^2}{f_I^2 \vevnoabs{\theta^2}} \right).
\ee
As such, employing \Eref{eq:axionfraction},
\be \label{eqn:fIonfa}
\frac{f_I}{f_a} \gsim 1.4 \times 10^4 \sqrt{R_a \left(\frac{r}{.002}\right) \left(\frac{.0019}{\alpha_\text{iso,max}}\right)} \left(\frac{10^{-1} M_P}{f_a}\right)^{5/12}
\ee
(valid for $f_a<\hat{f_a}$, while a similar expression can be derived for the other case).
Thus, for a theory with detectable gravitational waves $r \gsim 2 \times 10^{-3}$, a large separation is required between $f_I$ and $f_a$ to evade bounds on isocurvature.

This displacement must be generated in a way consistent with inflation.
For instance, $\vev{S}$ could be displaced via a model that leverages inflationary dynamics to give an explicit inflationary minimum for $\abs{S}$ larger than its present one.  However, as seen from \Eref{eqn:fIonfa}, $f_I$ must be super-Planckian for $f_a \gsim 4 \times 10^{-7} M_P$.  Such a large displacement is likely to seriously disrupt the inflaton potential via the couplings between the inflationary and PQ sectors, making such a mechanism unfavorable; further details are provided in footnote \ref{fn:superPlanckian}.
As such, the likely source of the inflationary radial field displacement is Hubble friction, \ie, the situation in which the radial field is displaced from the minimum today at the onset of inflation and remains light (compared to $H_I$) during inflation.

In fact, regardless of the source of the radial field displacement, such super-Planckian field values could still disrupt inflation.  Depending on the form of the potential for $S$, at large field values it may dominate the Universe's energy density, superseding the supposed inflationary sector. This effect is exacerbated as the required $f_{I}$ increases.
Moreover, such large values for $f$ have resisted embedding in string theory \cite{Banks:2003sx}, perhaps related to the weak-gravity conjecture \cite{ArkaniHamed:2006dz,Brown:2015iha,*Brown:2015lia,Heidenreich:2015wga,delaFuente:2014aca}. 

This concern could be avoided if the PQ sector were responsible for inflation, as suggested in \cite{Fairbairn:2014zta}.  We will return to this example at the end of this section, but, as we shall see, the other constraints outlined below generally require additional late-time dilution of axions for such ``PQ sector inflation" to be a viable solution.

Following inflation, once $H$ decreases below the radial field's mass, $\sigma$ will start to oscillate with a large initial amplitude corresponding approximately to its displacement during inflation.\footnote{Adiabatic relaxation as described in \Sref{sec:adiabatic}, which requires an inflationary minimum set by inflationary dynamics, is not possible in this model which relies purely on Hubble friction.}  
As we discuss in \Sref{sec:additionalconstraints}, the amplitude and decay of these oscillations are constrained by cosmological observations---the energy density must be dissipated efficiently and appropriately (\Sref{sec:radialenergydensity}) and fluctuations induced in a light, Hubble-trapped $\sigma$ may also be bounded (\Sref{sec:radialperturbations}).
Both of these constraints can be evaded via appropriate model building that is largely independent of the axion.  Alternately, if $\sigma$ is the inflaton, these constraints are superseded by requiring generation of  the observed curvature perturbations and appropriate reheating.

Our main focus here, however, is that the axion field itself can experience nonperturbative effects from large $\sigma$ oscillations, 
along the lines of the parametric resonance effects that can be induced by inflaton oscillations during preheating \cite{Kofman:1994rk,Kofman:1997yn}.  Constraints from these effects turn out to be most relevant for smaller $f_{a}$, and thus complementary to the concerns about the disruption of the inflation potential.  Specifically, radial oscillations can excite fluctuations in the axial field via parametric resonance.  If the fluctuations in the axial field grow to order $f_a$ so that $\theta=a/f_a$ takes all possible values throughout the universe with roughly equal probability, topological defects form and symmetry can be considered restored \cite{Tkachev:1998dc}.\footnote{The fluctuations can also be viewed as inducing a large, positive effective mass squared for the radial direction such that the potential no longer exhibits a symmetry-breaking form \cite{Tkachev:1998dc}.}
Since topological defects overclose the universe for the models with larger $f_a$ (or $N_{\rm DW}$) that are of interest here, PQ symmetry restoration must be avoided.\footnote{In addition to the possibility of symmetry restoration, large local fluctuations could contribute to the axion relic density (though not isocurvature, as the relevant modes are very small wavelength) or even result in the field locally relaxing to different minima after inflation, producing domain walls. As fluctuations grow exponentially, though, these latter concerns are only likely to be relevant very close to the region where there is risk of fully restoring the symmetry.}
Moreover, if the symmetry is restored, this clearly defeats the purpose of suppressing the axion fluctuations in the first place.

Axion fluctuations are more strongly enhanced for larger initial $\sigma$ oscillation amplitude, which results in an increased duration of oscillations and oscillation speed.
Based on lattice simulations \cite{Kawasaki:2013iha}, topological defects form when the initial amplitude
\be \label{eq:symrestoration}
\abs{S}_i \gsim 10^4 \left(\frac{f_a}{\sqrt{2}}\right)
\ee
in a quartic potential and matter-dominated background.\footnote{Refs.~\cite{Tkachev:1998dc, Kasuya:1999hy, Kawasaki:2013iha} use the linear basis, $S=\frac{1}{\sqrt{2}}(X+iY)$, to explore this effect, but it can also be understood in the nonlinear basis of \Eref{eqn:Snonlin} via derivative interactions between $\sigma$ and $a$, see \cite{Mazumdar:2015pta}.}
The bound is even stronger in the case where the PQ field dominates the energy density \cite{Tkachev:1998dc, Kasuya:1999hy}, in which case topological defects form when
\be \label{eq:symrestPQdom}
\abs{S}_i \gsim 2 \times 10^2 \left(\frac{f_a}{\sqrt{2}}\right).
\ee
A similar bound should apply in the case of radiation domination as a PQ field oscillating in a quartic potential behaves as radiation.
Thus, avoiding symmetry restoration requires that the initial amplitude of oscillations (or, at least, the initial amplitude when the potential starts behaving quartically \cite{Harigaya:2015hha}) is somewhat small.\footnote{Another more model-dependent nonthermal effect that can restore the PQ symmetry is described in \Sref{sec:scalartrapping}.}  

Comparing \Erefs{eqn:fIonfa}{eq:symrestoration}, for $f_a \lsim 10^{-1} M_P$, one cannot simultaneously suppress isocurvature using wave function renormalization alone and avoid symmetry restoration. Note $f_I$ in \Eref{eqn:fIonfa} is the field value when the modes relevant to the CMB exit the horizon, as opposed to when oscillations commence. However, we expect $\abs{S}_i$ to differ from $\frac{f_I}{\sqrt{2}}$ only by ${\cal O}(10\%)$ due to the flatness of the PQ potential required to avoid dominating the inflationary energy density. Thus, suppressing isocurvature via this mechanism is challenging and likely only viable for a limited range of $f_a$ and potentials.

The constraint from nonthermal symmetry restoration could be relaxed somewhat if the PQ field began oscillating in a potential dominated by $\abs{S}^{2 M}$ with $M \geq 3$ \cite{Harigaya:2015hha}.  However, constraints still apply once quartic or quadratic terms come to dominate.
Alternatively, the required hierarchy in $f_I$ and $f_a$ could be reduced in the case of a late-decaying scalar that dilutes the axion relic abundance \cite{Steinhardt:1983ia,Kawasaki:1995vt,Fox:2004kb}.  In this case,
\be
\label{eq:axionfractiondiluted}
R_a \simeq 1.7 \times 10^7 \left(\frac{\trh}{6~\mev}\right) \vevnoabs{\theta^2} \left(\frac{f_a}{M_P}\right)^2,
\ee
where $\trh$ is the reheat temperature at which the scalar decays, with the requirements $\trh \gsim 6~\mev$ to not spoil Big Bang nucleosynthesis (BBN) \cite{Reno:1987qw} and $\trh \lsim \Lambda_{\rm QCD}$ for the dilution to occur.  Then, the isocurvature bound requires
\be \label{eq:fIonfadiluted}
\frac{f_I}{f_a} \gsim 3.0 \times 10^3 \sqrt{R_a \left(\frac{r}{.002}\right) \left(\frac{.0019}{\alpha_\text{iso,max}}\right) \left(\frac{\trh}{6~\mev}\right)},
\ee
which could still be permitted by \Eref{eq:symrestoration} depending on $r$ and $\alpha_{\rm iso}$.
However, future nonobservation of isocurvature would ultimately restrict even this case.

Finally, even if axion fluctuations are made negligible, the initial displacement of the axion field is also constrained to avoid the overproduction of  DM, particularly for larger values of $f_a$.
\Eref{eq:axionfraction} implies
\be
\label{eq:thetai}
\theta_i \lsim \left\{
\begin{array}{ll}
1.3 \times 10^{-4} (f_a/M_P)^{-7/12}, & f_a<\hat{f}_a, \\
7.9 \times 10^{-5} (f_a/M_P)^{-3/4}, & f_a>\hat{f}_a.
\end{array}
\right.
\ee
In other words, the inflationary value for the PQ phase must not differ significantly from the value today.
Here, the initial misalignment at the onset of inflation corresponds to that in the Hubble patch that gave rise to our Universe.
Such a small misalignment angle could be justified by an anthropic argument---the axion takes on different initial misalignments in different inflationary patches, and we reside in one such patch that gives rise to a not-too-large axion abundance \cite{Linde:1987bx,Wilczek:2004cr}.
Alternatively, dilution of the axion abundance by a late-decaying particle would permit larger initial misalignment angles for a given $f_a$.

Overall, it seems challenging to implement a solution in which axion isocurvature is suppressed by wave function renormalization alone.  

\subsection*{PQ Sector Inflation}

Some of these challenges, namely those arising from the interplay between the PQ and inflationary sectors, could be surmounted if these sectors were identified.
However, this approach is complicated by nonperturbative effects that result in the PQ symmetry being restored in much of the same parameter space where the isocurvature fluctuations are sufficiently suppressed.  As we shall see, this leads to the additional requirement of late-time dilution of the axion abundance to yield a viable inflationary model.

If the PQ field is in a ``wine-bottle'' potential,
\be \label{eqn:winebottle}
V=\lambda \left(\abs{S}^2-\frac{f_a^2}{2}\right)^2,
\ee
it will have a $\abs{S}^4$-type inflationary potential \cite{Linde:1983gd}, which is inconsistent with BICEP2/Keck/Planck data on $r$  and the scalar tilt $n_{s}$ \cite{Ade:2015xua,Array:2015xqh,Ade:2015lrj}.
The introduction of a nonminimal coupling of $S$ to gravity \cite{Spokoiny:1984bd}
\be
\label{eq:nonmincoupling}
V \supset \xi \abs{S}^2 R,
\ee
where $R$ is the Ricci scalar and $\xi$ is a constant, allows for an inflationary model potentially consistent with existing data. 
The measured value of $\Delta_R^2$ specifies the relationship between $\xi$ and $\lambda$.  With the potential so fixed, the slow-roll parameters and inflaton field value throughout inflation (and thus $f_I$) can be calculated.  One finds the requirement $\xi \gsim (\text{few}) \times 10^{-3}$ to satisfy bounds on $r$ and  $n_s$.  The details are worked out in Ref.~\cite{Fairbairn:2014zta} (see also \cite{Bezrukov:2007ep, *Bezrukov:2008dt}), which identifies a window $10^{12}~\gev \lsim f_a \lsim 10^{15}~\gev$ where isocurvature constraints can be avoided.

However, this model suffers from the previously described parametric resonance constraints, wherein oscillations in $\sigma$ can restore the symmetry. Because the PQ field dominates the energy density, a stronger bound more analogous to \Eref{eq:symrestPQdom} applies. However, due to the nonminimal coupling, we cannot directly apply the bound of \Eref{eq:symrestPQdom}.  First, the potential is modified: in the Einstein frame,
\be
V_E=\frac{\lambda(\abs{S}^2-f_a^2/2)^2}{(1+2 \xi \abs{S}^2/M_P^2)^2},
\ee
which for large values of $\abs{S}$ and $\xi > 0$ is flatter than the wine bottle.  Second, 
$\sigma$ has a rolling mass parametrized by,
\be
Z(\sigma)=\frac{(1+\xi \sigma^2)^2}{1+\xi \sigma^2 + 6 \xi^2 \sigma^2},
\ee
where a heavier rolling mass, $Z(\sigma)<1$, corresponds to slower motion.  

Rather than attempting a full reanalysis similar to that of \cite{Tkachev:1998dc, Kasuya:1999hy} including these contributions, we present two bounds that we believe bracket the true one. They both come from the rule of thumb of \Eref{eq:symrestPQdom}, but use different choices of $\abs{S}_i$. In the first, $\abs{S}_i \simeq \abs{S}_{\text{end}}$, where $\abs{S}_{\text{end}}$ is the value for $\abs{S}$ at the end of inflation, defined to be when the slow-roll parameter $\epsilon \simeq 1$. This requirement is likely a bit too restrictive, especially for larger values of $\xi$, since the radial field will begin rolling more slowly due both to its rolling mass and the shallower potential compared to the case of $\xi=0$.  The second possibility is to take $\abs{S}_i$ to be the point after inflation has ended where both $Z(\sigma) \simeq 1$ and $\frac{d \log V_E}{d \log \sigma} \simeq 4$, \ie, where the rolling mass is close to its standard value of 1 and the potential is similar to \Eref{eqn:winebottle}.  This second bound is likely not quite constraining enough, as the radial field will already have begun rolling from a larger field value and will have attained some ``velocity'' from doing so.

\begin{figure}
\includegraphics[width=\columnwidth]{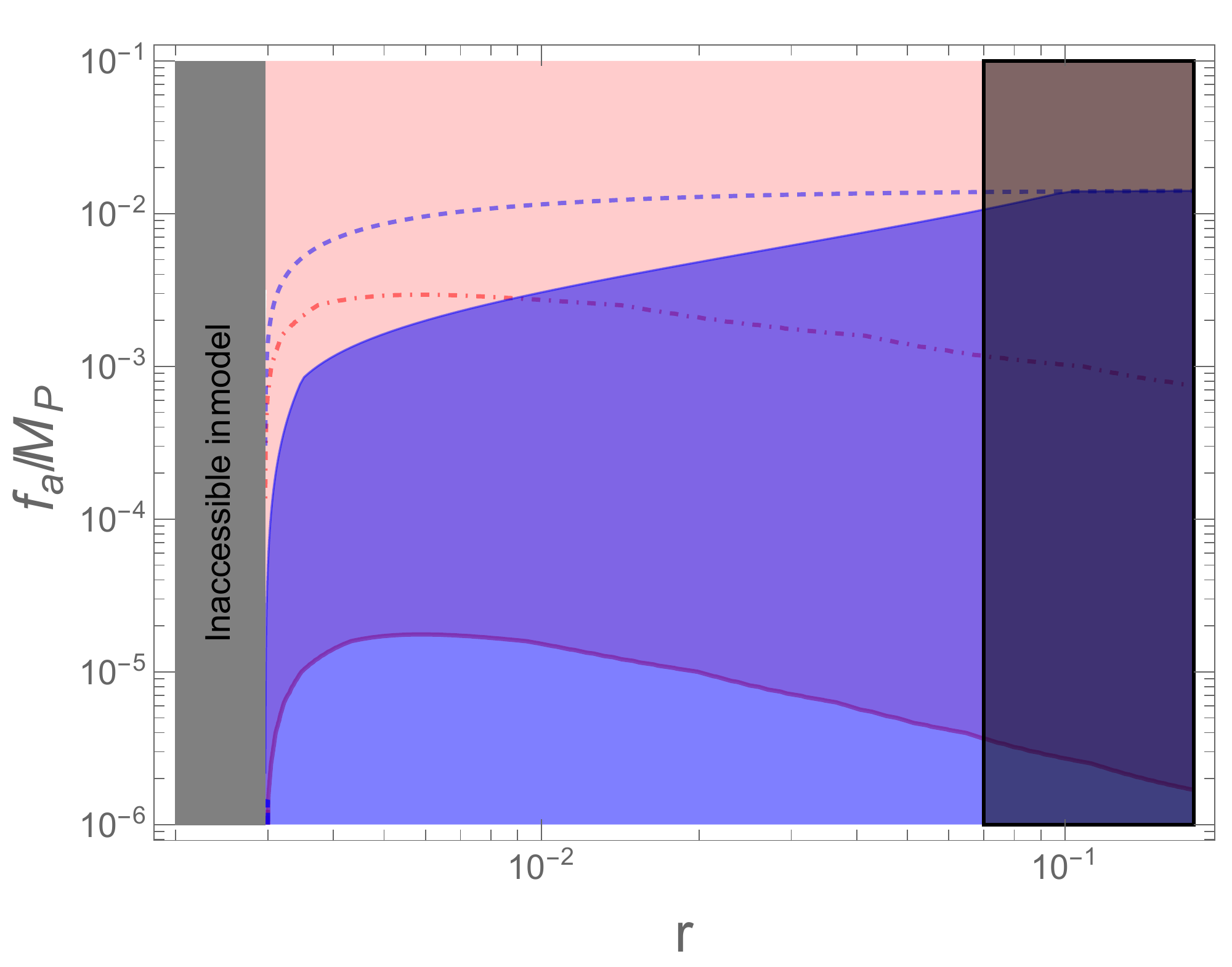}
\caption{\label{fig:PQFieldInflation} Constraints on the radial part of the PQ field acting as the inflaton.  The red curves denote isocurvature constraints after taking into account $f_I>f_a$ assuming $R_a=1$ with an abundance determined by either \Eref{eq:axionfraction} (solid, shaded upwards) or \Eref{eq:axionfractiondiluted} with $\trh=6~\mev$ (dot-dashed).  The two blue curves show two prescriptions for determining where the PQ symmetry may be restored by parametric resonance (see text for details); the solid shaded bound should be viewed as definitively excluded.
The black vertical region on the right shows the current BICEP2/Keck/Planck bounds on the tensor to scalar ratio, and the gray vertical region on the left is inaccessible in this model.}
\end{figure}

These symmetry restoration bounds are plotted in \Fref{fig:PQFieldInflation}, with the solid blue curve and shaded region corresponding to constraints with $\abs{S}_i$ determined by conditions on $Z(\sigma)$ and $V_E$,\footnote{In our study, we require $Z \geq 0.9$ and $\frac{d \log V_E}{d \log \sigma} \geq 3.9$.  The solid blue curve is mildly sensitive to this choice, but the overlap of the regions excluded by the solid blue and red curves is not.} and the dashed blue curve corresponding to taking $\abs{S}_i = \abs{S}_{\rm end}$.  Also shown  are isocurvature bounds if the axion makes up all of the DM abundance (\ie, $R_a=1$) assuming either \Eref{eq:axionfraction} (solid red region)\footnote{Our bounds on isocurvature are stronger by roughly a factor of $8 \pi$ compared to Ref.~\cite{Fairbairn:2014zta} because their Eq.~(15) uses the Planck mass while their other expressions use the reduced Planck mass.  We also use the updated Planck bounds on $\alpha_\text{iso}$ \cite{Ade:2015xua}.} or \Eref{eq:axionfractiondiluted} with maximal possible dilution (corresponding to $\trh=6~\mev$) (dot-dashed red line).
For the purposes of calculating the isocurvature constraints we take $f_I$ to be the field value 60 e-folds before the end of inflation, approximately corresponding to the scale when modes relevant to the CMB exit the horizon.
Also shown are current BICEP2/Keck/Planck bounds on $r$ \cite{Array:2015xqh,Ade:2015lrj}.  Note that future experiments will be able to probe all values of $r$ for this model \cite{Creminelli:2015oda}.

Comparing even the conservative estimate of the symmetry restoration bounds to the isocurvature bounds assuming the axions are produced in the manner of \Eref{eq:axionfraction} (\ie, without a late-decaying scalar), it is clear that some additional cosmological model building is necessary to make this a viable inflationary model.  With late-time dilution, some parameter space may not yet be excluded depending on where the true symmetry restoration bounds fall.  Nevertheless, this parameter space is narrow and only potentially allows PQ-breaking scales $f_a \lsim 2 \times 10^{-3} M_P$.  Additionally, a detection of $r \gsim 10^{-2}$ would definitively exclude this model.  Of course, larger values of $f_a$ and $r$ could be allowed if the requirement that $R_a=1$ is relaxed, though this would require an even smaller initial misalignment angle without any particular motivation for such a tuning.

This model illustrates an important tension. Sufficient suppression of isocurvature requires a very large hierarchy between $f_{I}$ and $f_{a}$. For just such a hierarchy, nonperturbative dynamics have the potential to restore the PQ symmetry. Consequently, additional cosmological mechanisms are necessary to permit models that suppress isocurvature via wave function renormalization, and the allowed values for $f_a$ are still restricted.  Moreover, if isocurvature perturbations continue unobserved, even these remaining values for $f_a$ may eventually be excluded.

Having demonstrated the difficulties inherent in models that seek to suppress isocurvature via axion wave function renormalization alone, we now turn to models in which field displacement enhances explicit PQ breaking in the early Universe.

 
\section{Axion Mass from Enhanced Explicit $U(1)_{\rm PQ}$ Breaking During Inflation}
\label{sec:axionmass}

Scalar field fluctuations are very efficiently suppressed for fields with large masses $m \gsim H_I$ (see, \eg, \cite{Riotto:2002yw}).
With this motivation, in this section we will consider models in which 
$U(1)_{\rm PQ}$ is only an approximate symmetry that is explicitly broken to a discrete subgroup $\mathbb{Z}_N$.  
Then, higher dimension operators allowed by the discrete symmetry but forbidden by the continuous $U(1)_{\rm PQ}$  generate contributions to the axion mass in addition to those from QCD.
If these operators were enhanced during inflation, the large inflationary mass could sufficiently suppress axion isocurvature while still yielding a model consistent with strong CP constraints.

Supposing the axial component of the field is heavy during inflation, 
it will evolve to the minimum favored by the operator(s) responsible for generating its large mass.
This minimum need not coincide with those of the QCD potential and in general would not be expected to as the two contributions to the axion potential arise from different sources.
However, it will determine the initial misalignment angle $\theta_i$.
So, for axions with large $f_a$, barring a mechanism for the axion to adiabatically evolve to the vicinity of the minimum today, cosmological considerations do require a rather dissatisfying coincidence between operators to avoid overproduction of axion DM, see \Eref{eq:thetai}.\footnote{The anthropic argument that we simply reside in an inflationary patch where the initial misalignment is small is no longer valid. Any attempted anthropic argument must instead consider a multiverse selection amongst  appropriate potential parameters.}
This incredible coincidence could be mitigated if, for instance, a late-decaying particle diluted the axion abundance, allowing larger initial misalignment.

Moreover, if non-QCD contributions to the axion potential are incompletely turned off, measurements of CP-violating observables today tightly constrain the size of such contributions \cite{Kamionkowski:1992mf}.
For the axion to still provide a solution to the strong CP problem, the minimum today must correspond to the QCD minimum $\theta_0$ to a very high degree,
\be
\vev{\frac{a}{f_a} - \theta_0} \leq \bar{\theta} \simeq 10^{-11},
\ee
where $\bar{\theta}$ represents the current constraints on the effective $\theta$ angle.  Let $m_{\rm eff, 0}^2$ represent additional contributions to $m_{a}^2$ today and $\theta_N$ represent the minimum favored by the additional contributions (such that $\theta_i \simeq \abs{\theta_N - \theta_0}$ is the initial displacement).
Then, assuming $\theta_i$ is tuned to be as small as required by cosmology but not more so,
\be
\label{eq:simplemeffconstraint}
m_{\rm eff, 0}^2 \lsim \frac{\bar{\theta} m_{\rm QCD}^2}{\abs{\theta_N - \theta_0}}.
\ee
Note, owing to the necessarily small denominator, this constraint is somewhat weaker than the usual constraint for ${\cal O}(1)$ displacement between the operators---constraints would be more stringent for larger displacements as allowed by, \eg, dilution.  Here we have approximated the QCD potential as quadratic in the vicinity of the minimum,
\be
\label{eq:VQCD}
V \simeq \frac{1}{2} m_{\rm QCD}^2 \left(a - f_a \theta_0\right)^2,
\ee
and $m_{\rm QCD} \simeq 6~\mu {\rm eV} \left(10^{12}~\gev/f_a\right)$ \cite{Agashe:2014kda}.

This severe constraint makes it impossible to arrange for a large inflationary mass via explicit PQ symmetry breaking under the assumption that $f_I = f_a$. However, displacement of the PQ field during inflation from its minimum today can enhance the non-QCD contributions to $m_a^2$.  In fact, these models can achieve an effective inflationary axion mass $m_{\rm eff, inf}^2 \gsim H_I^2$ even for $f_I \lsim M_P$, so the required field displacement can be substantially less than the (super-)Planckian values of $\vev{S}$ needed to suppress isocurvature by changing the normalization of the axion field alone (\Sref{sec:wfsuppression}).

However, the symmetry breaking responsible for giving the axion a large inflationary mass will also give the radial field a mass of the same order, $m_\sigma \sim H_I$.
Thus, in this case, $f_I  > f_a$ cannot simply arise from Hubble trapping, but instead requires a modification of the PQ potential during inflation.  As such, the PQ field must couple to inflationary dynamics.  
Importantly, such couplings to the inflationary sector can disrupt the flatness of the fragile inflaton potential.  
To be concrete, the contributions of couplings between the PQ and inflationary sectors, denoted $\Delta V$, will result in additional contributions to the slow-roll parameters,
\begin{align}
\label{eq:slowrollparams}
\epsilon & \equiv \frac{M_P^2}{2} \left(\frac{V'}{V}\right)^2, & \eta & \equiv M_P^2 \frac{V''}{V},
\end{align}
where primes denote derivatives with respect to $I$.
There are no hard and fast constraints on such contributions, as the part of the potential that depends only on the inflaton $V_I$ could always be tuned to give values consistent with CMB observations.
However, large contributions from $\Delta V$ would require particularly severe (and potentially dynamical, if $f_I$ changes over the course of inflation) tunings.  
As such, $V_I$ would have to be very carefully arranged to appropriately balance the contribution from $\Delta V$ throughout inflation and maintain an appropriate inflationary trajectory.

We study the restrictions arising from these contributions to the inflationary potential as well as from strong CP constraints in the context of specific models below.
We will consider three models.  The first will consist of only a single PQ field $S$ charged under $\mathbb{Z}_N$, and we will see that the combination of strong CP constraints and avoiding excessive disruption of the inflaton potential limits the values of $f_a$ that can be reasonably allowed.
Consequently, we will consider other models in which additional fields charged under $U(1)_{\rm PQ}$ (or the discrete symmetry)---either the inflaton or additional PQ fields---acquire large values during inflation, potentially relaxing constraints on models with larger $f_a$.

\subsection{A Simple Model with a Single PQ Field}
\label{sec:expbreak}

Perhaps the simplest example of an operator that explicitly breaks $U(1)_{\rm PQ}$ to a discrete symmetry (in this case, $\mathbb{Z}_N$) is $S^N$.  
We consider a basic model containing this operator as well as $U(1)_{\rm PQ}$ invariant terms,
\be
\label{eq:Sinfcoupling}
V \supset \lambda \left(\abs{S}^2 - \frac{f_a^2}{2}\right)^2 - \frac{\delta}{2} I^2 \abs{S}^2 + \left(\frac{k S^N}{M_P^{N-4}} + \hc\right).
\ee
The inclusion of the $\delta$ coupling is motivated by the fact that it is not forbidden by any symmetries.  In addition, for $\delta > 0$, this coupling can be responsible for generating the necessary radial displacement $f_I > f_a$.

The operator in the last term of this equation will generate an additional contribution to the axion mass
\be
m_{\rm eff}^2 = \frac{\abs{k} N^2 \vev{S}^{N-2}}{M_P^{N-4}}.
\ee
If $\theta_N$ represents the minimum favored by the $S^N$ operator,
strong CP constraints require the contribution to the axion mass today,
\be
\frac{\frac{m_{\rm eff, 0}^2}{N} \sin\left(N \abs{\theta_N - \theta_0}\right)}{m_{\rm QCD}^2 + m_{\rm eff, 0}^2 \cos\left(N \abs{\theta_N - \theta_0}\right)}
\lsim \bar{\theta},
\ee
where as above we have approximated the QCD potential as quadratic in the vicinity of the minimum, see \Eref{eq:VQCD}.  For small $m_{\rm eff, 0}^2$ (\ie, supposing $\sin\left(N \abs{\theta_N - \theta_0}\right)$ not incredibly small and so neglecting the subdominant term in the denominator), we write the constraint as
\be
m_{\rm eff, 0}^2 \lsim \frac{N \bar{\theta} m_{\rm QCD}^2}{\sin\left(N \abs{\theta_N - \theta_0}\right)}.
\ee
Taking $\theta_i = \abs{\theta_N - \theta_0}$ small as required by cosmology, see \Eref{eq:thetai}, and expanding this equation reproduces \Eref{eq:simplemeffconstraint}.  Unless $\sin(N\abs{\theta_N - \theta_0})$ is extremely close to zero---in particular, much closer even than required by cosmology---this places a stringent lower bound on $N$ for a given $(\abs{k}, f_a)$ that generally requires $N$ to be large, $N \gsim {\cal O}(10)$.  In other words, as one might expect, $U(1)_{\rm PQ}$ needs to be a good symmetry to a high degree in order to solve the strong CP problem.

Clearly, this precludes the $S^N$ operator from giving a large mass to the axion during inflation if $f_I = f_a$, even for $\abs{\theta_N - \theta_0} \ll 1$.
When $I$ takes on large values, though, the $I^2 \abs{S}^2$ coupling can drive $\vev{S} = \frac{f_I}{\sqrt{2}} > \frac{f_a}{\sqrt{2}}$.
Then, a large effective mass for the axial direction due to the $S^N$ term may stabilize the phase of the PQ field at $\theta_N$ and suppress fluctuations.  After inflation, the effect of the inflaton-PQ field cross-coupling will disappear, and the field will evolve to a minimum with $\vev{S} = \frac{f_a}{\sqrt{2}}$ and $\arg(S) \simeq \theta_N$ after a period of rolling and oscillation,
where it will remain until $m_{\rm QCD}^2$ turns on.\footnote{While the potential barriers in the axial direction are small at small $\vev{S}$, the larger barrier at large $\vev{S}$ will produce a ``funnel'' directing the phase towards $\theta_N$ or, if oscillations are sufficiently large to take $S$ through $\vev{S} = 0$, towards $\theta_N + \pi$. } 

However, this coupling also influences the inflaton potential.  The inflationary trajectory is constrained by limits on the slow-roll parameters given in \Eref{eq:slowrollparams},
\begin{align}
\label{eqn:rbound}
r & \approx 16 \epsilon < .07, \\
n_s - 1 & \approx - 6 \epsilon + 2 \eta \in (-0.0413,-0.0253),
\end{align}
where we have given the 95\% confidence bounds from \cite{Ade:2015xua,Array:2015xqh}.  A reasonable constraint to avoid an overly tuned inflationary model is to require that the additional contributions to $\epsilon, \eta$ do not exceed the maximal values consistent with present CMB observations.\footnote{An analogous constraint on $\xi$ or $\abs{\Delta V'''/V}$ from the running spectral index $\alpha_s$ is subdominant.  In addition, we have confirmed that $\Delta V \ll V_I$ in these models.}
For instance, requiring that the contribution be~ $\lsim 1$ of the maximal value corresponds to
\begin{align} \label{eqn:SNVpbound}
\abs{\frac{\Delta V'}{V}} & \leq 0.094, & \abs{\frac{\Delta V''}{V}} & \leq 0.021.
\end{align}
We stress though that, while it represents a severe, \emph{ad hoc}, dynamical tuning and so should be taken seriously, this constraint is aesthetic rather than experimental.\footnote{\label{fn:superPlanckian}This constraint generally precludes the $\vev{S} > M_P$ required in \Sref{sec:wfsuppression} from arising in this manner.  For instance, supposing $\Delta V = - \frac{\delta}{2} I^2 \abs{S}^2$ we expect $\delta I^2 \gsim H_I^2$ such that $m_\sigma^2 \gsim H_I^2$ and $\abs{S}$ evolves to the large vev.  This implies $\abs{\Delta V'/V} \sim f_I^2/(I M_P) \gsim 1$ and $\abs{\Delta V''/V} \sim f_I^2/I^2 \gsim 1$, exceeding the stated bounds.}

\begin{figure}
\includegraphics[width=\columnwidth]{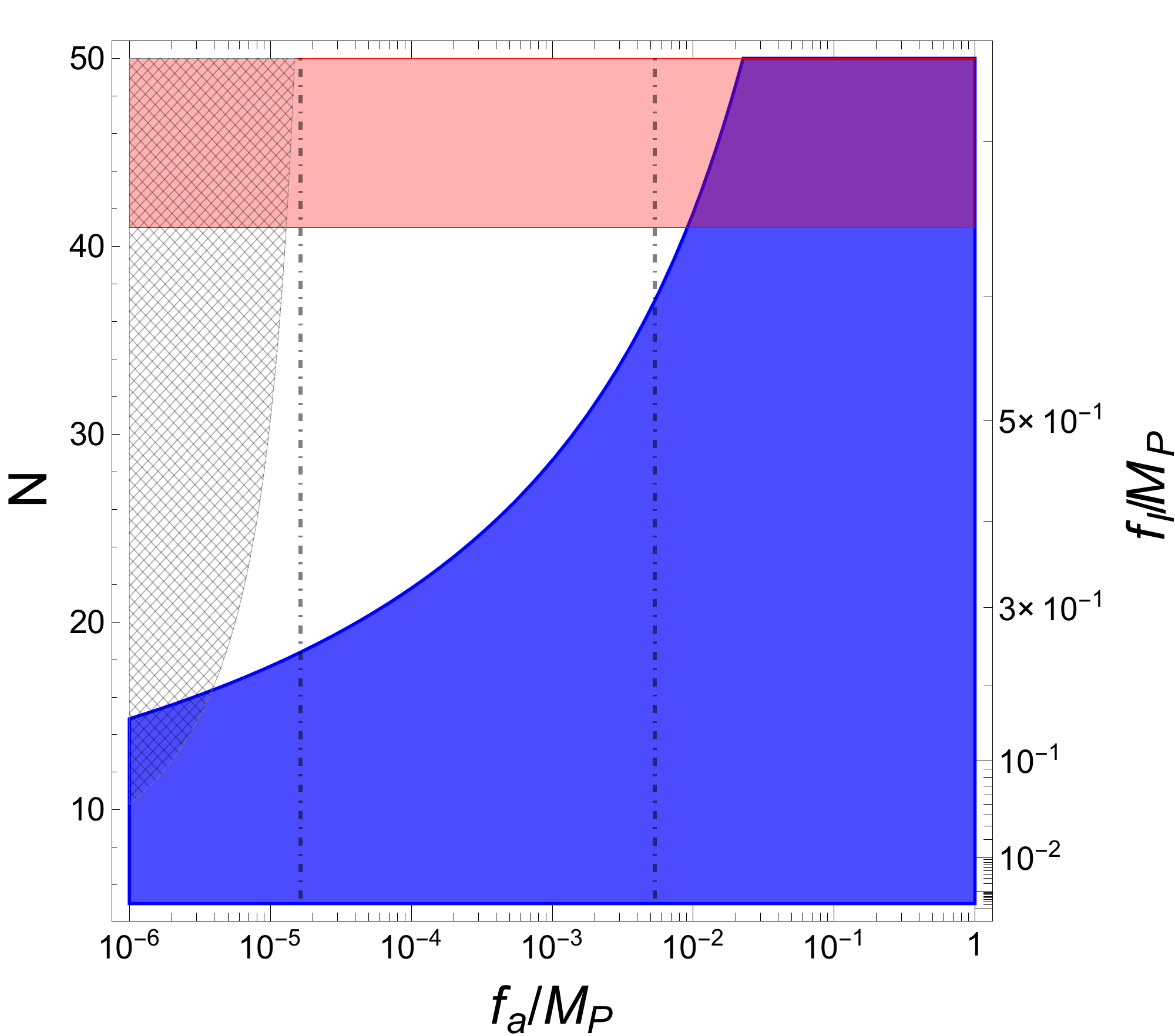}
\caption{\label{fig:SNconstraints} Bounds on the $S^N$ model in the parameter space of $f_a$ and $N$ for $\abs{k}=1$.  The right axis shows the required value for $f_I$ to give the axion a large enough inflationary mass to suppress isocurvature.  Blue corresponds to the strong CP constraint.  
Red indicates where the contributions to the slow-roll parameters from the $S$ potential are greater than present bounds.  The hatched region indicates where $\vev{S}_e>10^4 \frac{f_a}{\sqrt{2}}$ so PQ symmetry restoration via parametric resonance may occur (see text for caveats).  Vertical lines indicate the CASPEr Phase 2 (right) and ideal (left) reach (prospective bounds extend to the right).}
\end{figure}

In \Fref{fig:SNconstraints}, we present an example of the constraints on such a model.  Blue regions are excluded by strong CP constraints with cosmological bounds on the axion abundance taken into account, \ie, $\abs{\theta_N - \theta_0}$ set by \Eref{eq:thetai}.
The constraints are indiscernibly more stringent if one assumes $\abs{\theta_N - \theta_0} \simeq {\cal O}(1)$ as might be expected in the case of late-time dilution.
The red region denotes where the inflaton potential is significantly modified, in violation of \Eref{eqn:SNVpbound}.
If these constraints are weakened by a factor of $3$, this exclusion region moves up to $N \gsim 130$, potentially allowing $f_a \simeq 3 \times 10^{-1}$.
We take $H_I \simeq 10^{-5} M_P$, consistent with $10^{-3} \lsim \, r \, \lsim 10^{-2}$ that will be probed in near future experiments.\footnote{Note that $f_I > \frac{H_I}{2\pi}$ throughout the parameter space, such that the PQ symmetry is broken during inflation even for $f_a < \frac{H_I}{2\pi}$.}
The Lyth bound \cite{Lyth:1996im} implies that, for such values of $r$, the excursion of the inflaton field over the course of inflation is $\frac{\Delta I}{M_P} \gsim 1$.  So, we set $I_\ast = 5 M_P$ at CMB mode horizon crossing (comparable to the value for a Starobinsky-like model at the edge of observability \cite{Starobinsky:1980te}).

The constraints shown are conservative in the sense that, for each value of $N$, we choose $\lambda, \delta$ to be the smallest possible values such that the radial and axial directions both have masses $\gsim H_I$---larger couplings would result in larger contributions to the slow-roll parameters.\footnote{The required values of $\delta$ are sufficiently small that parametric excitation of $S$ due to inflaton oscillations is not a concern.}
The axion mass condition is equivalent to requiring 
\be
f_I \simeq \left(\frac{H_I^2 M_P^{N-4}}{\abs{k} N^2}\right)^{N-2}.
\ee
The CASPEr experiment \cite{Budker:2013hfa}---with its Phase 2 reach of $f_a \gsim 1.3 \times 10^{16}~\gev$ and ideal reach of $f_a \gsim 4 \times 10^{13}~\gev$---may eventually be able to probe much of the allowed region as indicated by the vertical lines.

The cross-hatched region denotes where the field value at the end of inflation $\vev{S}_e > 10^4 \frac{f_a}{\sqrt{2}}$, supposing that the value of the inflaton field at the end of inflation $I_e$ is a factor of 5 smaller than at CMB horizon crossing (such that $\frac{\vev{S}_e}{f_I/\sqrt{2}} \approx \frac{1}{5}$).  According to the analysis of \cite{Kawasaki:2013iha}, for such values radial PQ field oscillations may excite large oscillations in the axial field, potentially leading to nonthermal symmetry restoration (see \Eref{eq:symrestoration} and surrounding discussion).  
In this model, the situation is complicated by the presence of additional operators and the lack of initial fluctuations.
For instance, the curvature of the potential in the axial direction due to the $S^N$ operator may suppress fluctuations.  Alternatively, self-couplings of the axial field resulting from this operator may mitigate the growth of fluctuations \cite{Prokopec:1996rr}.
Thus, while this region is not necessarily excluded, a further analysis of field dynamics post-inflation would be required to ensure a consistent cosmology.

This analysis suggests that it is difficult to reach larger values of $f_a \gsim 10^{-2} M_P$.  While the conclusion that a significant modification of the PQ potential can disrupt inflation is robust, the exact limits do depend on the inflationary model.
For larger values of $I_\ast$, the same $f_I$ can be achieved for smaller values of $\delta$ and $\lambda$, resulting in smaller contributions to slow-roll parameters.
In addition, as long as  $I_e \simeq M_P$, $\vev{S}_e$ would be reduced in the case of large $I_\ast$, shrinking the nonthermal symmetry restoration region.
As such, perhaps paradoxically, high-scale inflation may reconcile more readily with a solution of this type---the steeper potential could be less susceptible to disruption.
It should be noted, however, that larger $I_\ast$ would also likely correspond to larger values of $H_I \simeq 10^{-4} M_P$ ($r \simeq 10^{-2}$), implying imminent tensor mode observation.
Meanwhile, in a specific inflationary model, violation of \Eref{eqn:SNVpbound} may result in a worse than ${\cal O}(1)$ tuning.  Recall, these equations were derived using current experimental constraints.  For a Starobinsky-like model with $r \simeq 2 \times 10^{-3}$,  the red region in \Fref{fig:SNconstraints} would more approximately correspond to tuning $\gsim$ few, owing to the smaller denominators in \Eref{eqn:SNVpbound}.  Requiring tuning $\leq 1$ in this model would exclude $N \gsim 31$, $f_a \gsim 2 \times 10^{-3} M_P$.

In sum, strong CP constraints require large values of $N$ to ensure that the contribution to the axion mass due to the explicit breaking is small today.  As such, $f_I$ must be somewhat larger than $f_a$ to ensure the axion is sufficiently heavy during inflation.  In other words, large modifications to the PQ potential are required.  These in turn ``backreact'' on the inflaton potential, at worst threatening to destabilize the fragile inflaton potential and at best constituting a very severe (field-dependent) tuning.

One reason that it is difficult to achieve a sufficiently large mass whilst satisfying strong CP constraints is that the same field is responsible for both the enhancement of the explicit breaking and solving the strong CP problem---the latter requires the potential of this field be largely $U(1)_{\rm PQ}$ invariant today, such that drastic modifications of its potential are required during inflation, readily disrupting inflationary dynamics.  Consequently, in the next subsections, we discuss the possibility of realizing a large axion mass during inflation by coupling $S$ to additional fields that acquire large vevs during inflation but are not subject to the same strong CP constraints.  We will see that such approaches do indeed extend the reach in $f_a$ relative to this simple model.

\subsection{Inflaton-Sourced Axion Mass}
\label{sec:ISNmodel}

At first glance, coupling to the inflaton would appear an ideal method for boosting the axion mass during inflation---the large (super-Planckian) value of $I_\ast$ could give a significant contribution to $m_{\rm eff, inf}^2$ even via Planck-suppressed operators.  Meanwhile, for small $\vev{I}$ today, this contribution would be suppressed, alleviating strong CP constraints.  However, as we discuss, the difficulty in this approach lies in generating the desired operators without generating either additional unwanted operators or symmetries (the latter of which, for instance, may prevent fields from acquiring a necessary mass).

One straightforward approach is to consider a term that explicitly breaks $U(1)_{\rm PQ}$ such as
\be
\label{eq:expbreakISN}
V \supset \frac{k I S^N}{M_P^{N-3}} + \hc
\ee
Such an operator was considered in, \eg, \cite{Higaki:2014ooa}.\footnote{Refs.~\cite{Kadota:2014hpa,Kadota:2015uia} consider the potential for similar operators to generate a cross-correlation spectrum in the case where isocurvature is unsuppressed.}
If the rest of the potential maintains the symmetry $I \rightarrow -I, S^N \rightarrow -S^N$, this explicitly breaks $U(1)_{\rm PQ} \rightarrow \mathbb{Z}_{2N}$.  Then, for $\vev{I} = 0$, the leading operator expected to contribute to the axion mass would involve $S^{2N}$, permitting smaller values of $N$ for a given $(\abs{k}, f_a)$ while still maintaining a solution to the strong CP problem.

However, for $\vev{S} = \frac{f_a}{\sqrt{2}} \neq 0$, \Eref{eq:expbreakISN} generates a tadpole for $I$, leading to $\vev{I} \neq 0$ unless the potential is specifically tuned to stabilize $\vev{I} \approx 0$.  
If the inflaton potential is approximately quadratic near the origin, 
\be
V_I \simeq \frac{1}{2} m_I^2 I^2,
\ee
then,
\be
\label{eq:Ivev}
\vev{I} \simeq \frac{2 \abs{k} \left(f_a/\sqrt{2}\right)^N}{m_I^2 M_P^{N-3}} \cos\left(N \vev{\frac{a}{f_a} - \theta_N}\right).
\ee
This in turn drives $\vevnoabs{\frac{a}{f_a} - \theta_0} \neq 0$---in fact, due to additional $\left(\frac{M_P}{m_I}\right)^2$ enhancement, the effect of this term is expected to dominate over that of an $S^{2N}$ operator supposing both operators are generated with similarly sized coefficients (unless $m_I \sim M_P$).  Consequently, strong CP considerations will still imply a somewhat stringent bound on $N$ that is not quite a factor of $2$ weaker than that for the model considered in \Sref{sec:expbreak}.

\begin{figure}
\includegraphics[width=\columnwidth]{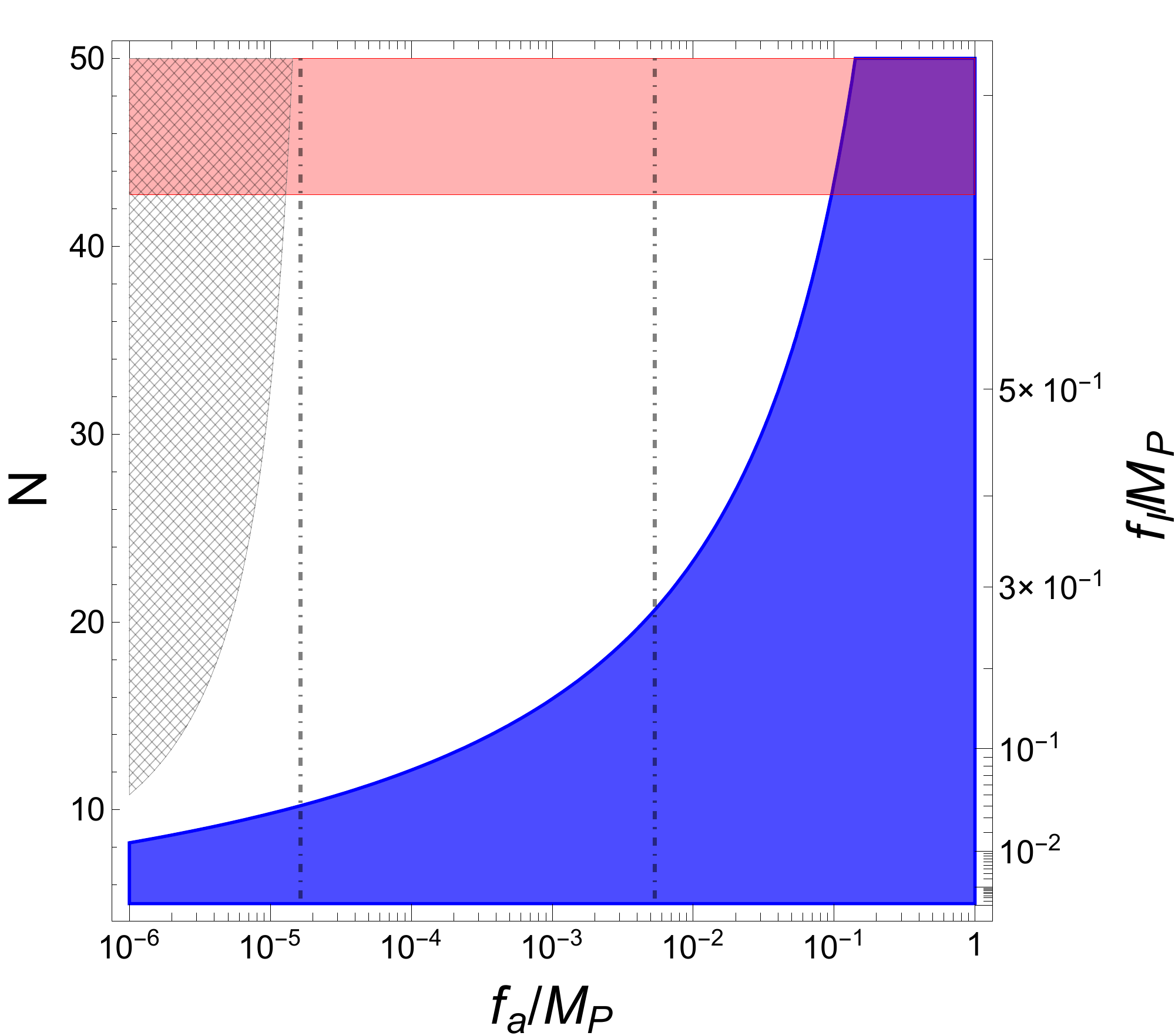}
\caption{\label{fig:ISNconstraints} Bounds on $IS^N$ model with $m_I=10^{-5} M_P$ and $\abs{k} = 1$.  All curves are the same as in \Fref{fig:SNconstraints}.}
\end{figure}

Regardless, the combination of lower $N$ and large $I_\ast$ does allow a sufficiently large axion mass to be achieved with a smaller increase of $\vev{S}$ during inflation, and hence less risk of destabilizing the inflaton potential.  This opens up some parameter space that was not available in the model of \Sref{sec:expbreak}.  In \Fref{fig:ISNconstraints}, we show the analog of \Fref{fig:SNconstraints} with the $S^N$ operator of \Eref{eq:Sinfcoupling} replaced by the $I S^N$ operator of \Eref{eq:expbreakISN} assuming $m_I \approx 10^{-5} M_P$ (representative of large-field models such as chaotic or Starobinsky-like inflation).  Again, caveats about inflation model dependence such as those in the previous subsection apply.

This analysis assumes that \Eref{eq:expbreakISN} is the leading contribution to the axion mass.  While operators with more powers of $S$ will be suppressed as $\vev{S} < M_P$, this is not the case for operators including higher powers of $I$.  For instance, the $I \rightarrow -I, S^N \rightarrow -S^N$ symmetry discussed above would allow operators of the form $I^{2M+1} S^N$.  Such operators may well be subdominant, though, as assumedly there exists some symmetry (for instance, a shift symmetry) responsible for maintaining the flatness of the inflaton potential.  Hence, all operators containing $I$ would also come with a spurion representing the breaking of this symmetry---if  the appropriate combination to consider were, for instance, $(kI)^{2M+1}$, (where $k$ is now the aforementioned spurion) then for $\abs{k} \ll \frac{I_\ast}{M_P}$ such higher-dimension operators would also be suppressed and the above analysis would hold.
Reducing $\abs{k}$ reduces both the inflationary mass (such that larger $f_I$ would be necessary to yield $m_a \gsim H_I$) and the inflaton vev today, \Eref{eq:Ivev}.  As such, the red and blue regions in \Fref{fig:ISNconstraints} would both move down, but by different amounts, resulting in a slightly reduced reach in $f_a$.

\subsubsection*{Challenges for Realizing $I^M S^N$ with $M > 1$}

A tadpole for $I$ would not arise if the leading explicit-breaking operator contained a higher power of $I$, $I^M S^N$.  But, in such a case, it is hard to imagine what (beyond coincidental cancellation) could effectively suppress lower dimension operators.

If $I$ is real, then for odd $M$ this operator maintains the $I \rightarrow -I, S^N \rightarrow -S^N$ symmetry that permits the operator $I S^N$.  For the reasons discussed above, we may well expect $I S^N$ to dominate.  Even less promising is the scenario in which $M$ is even, as then the explicit breaking term violates $S^N \rightarrow -S^N$ and so nothing should forbid the operator $S^N$.

If $I$ were a complex field, one could imagine a global $U(1)$ with charges $q_S = \frac{1}{N}$, $q_I = -\frac{1}{M}$ that, for appropriate choices of $N$ and $M$, would give rise to $I^M S^N$ without additional lower dimension operators.  However, this global $U(1)$ must be explicitly broken to avoid the presence of a massless field during inflation---in fact, it is indistinguishable from $U(1)_{\rm PQ}$.  In this setup, it becomes difficult to achieve a single, light inflaton (with mass $\ll H$) without additional light fields.  If this $U(1)$ is a good symmetry during inflation (\ie, broken sufficiently negligibly to suppress the generation of other worrisome operators), it should also relate the masses of the real and imaginary components of $I$ to a good degree.  Thus, if some component of $I$ is light, the orthogonal component would be as well, potentially leading to isocurvature.  Alternatively, if the $U(1)$ is sufficiently badly broken that one component of $I$ is quite massive while the orthogonal component is light (the inflaton), then it would also not be good enough to forbid other operators.

In supersymmetry (SUSY), holomorphy and nonrenormalization can suppress or forbid operators that one may have expected to be present from a more na\"ive symmetry analysis.
Thus, SUSY appears promising for generating the desired $I^M S^N$ operator without dangerous operators involving fewer powers of the inflaton field.
However, once SUSY and $U(1)_{\rm PQ}$ are broken, dangerous operators contributing to the axion mass will be generated.  So, while SUSY can ensure that these operators exhibit additional suppression relative to nonsupersymmetric models, for instance by the scale of SUSY breaking, this suppression is generally insufficient to circumvent the very stringent strong CP constraints.

Moreover, there is the well-known challenge that it can be difficult to realize large-field models of inflation within supergravity (SUGRA) (for a review, see, \eg, \cite{Yamaguchi:2011kg}).
The SUGRA scalar potential is given as a function of $K$ and the superpotential $W$ by
\be
\label{eq:SUGRAV}
V_F = e^{K/M_P^2} \left(D_{\Phi_i} W K_{ij^\dagger}^{-1} D_{\Phi_j^\dagger} W^\dagger - 3 \frac{\abs{W}^2}{M_P^2} \right),
\ee
with $F_{\Phi_i} \equiv D_{\Phi_i} W = W_i + \frac{W K_i}{M_P^2}$ where $W_i \equiv \frac{\partial W}{\partial \Phi_i}$ (and similarly for $K$).
The exponential factor $\exp(K/M_P^2)$ induces large curvature at super-Planckian field values, a challenge for ensuring slow roll.  Thus, a viable model must generally invoke a symmetry (such as a shift symmetry) to prevent the inflaton from appearing in $K$ \cite{Kawasaki:2000yn}.
In addition, supersymmetry implies that the field containing the inflaton is necessarily complex, so any continuous symmetries relating the various components of the field must be badly broken to ensure only one component remains light.  This breaking combined with the presence of operators $I^M S^N$ contributes to the breaking of $U(1)_{\rm PQ}$.

To illustrate these points, consider a SUSY model with the following K\"ahler and superpotentials
\begin{align}
K & = \sum_{\Phi = X, Y, S, \bar{S}} \Phi^\dagger \Phi + \frac{1}{2} (I + I^\dagger)^2, \\
W & = \lambda Y \left(S \bar{S} - \frac{f_a^2}{2}\right) + m_I X I + k X I \frac{S^N}{M_P^{N-1}}.
\end{align}
The K\"ahler potential respects a shift symmetry $I \rightarrow I + i c$ for $c \in \mathbb{R}$, which is softly broken by $m_I$ and $k$.  This precludes the imaginary component of $I$ from appearing in $\exp(K/M_P^2)$, giving rise to a viable model of chaotic inflation in which the imaginary component serves as the inflaton.\footnote{Models of chaotic inflation are currently disfavored by bounds on $r$ from Planck, so such a model would have to be modified to be consistent with experimental results.  This could be done by, for instance, adding a nonminimal curvature coupling for $I$ similar to \Eref{eq:nonmincoupling}.  However, as our focus is the difficulty of embedding explicit PQ breaking in SUSY, we refrain from constructing a complete inflationary model here.}  The fields $S$ and $\bar{S}$ carry opposite PQ charge and the PQ symmetry is spontaneously broken today by $\vevnoabs{S \bar{S}} = \frac{f_a^2}{2}$ due to the first term in $W$.  

Supposing  $X$ and $Y$ are stabilized at the origin, $X = Y = 0$, the SUGRA scalar potential is
\be
V = e^{K/M_P^2} \left\{\abs{\lambda}^2 \abs{S \bar{S} - \frac{f_a^2}{2}}^2 + \abs{m_I + k \frac{S^N}{M_P^{N-1}}}^2 \abs{I}^2\right\},
\ee
in which the leading inflaton-PQ coupling comes from $\abs{I}^2 S^N$, without an $S^N$-type term at this level.  While no symmetry could prevent the generation of a term in the K\"ahler potential such as
\be
\Delta K = \frac{k m_I^\dagger S^N}{M_P^{N-1}} + \hc,
\ee
the contribution to the axion mass would vanish for $I = 0$ and $\vevnoabs{S \bar{S}} = \frac{f_a^2}{2}$ today.  It would appear that the constrained form of the potential permitted in SUSY allows the desired inflationary explicit-breaking operators to be generated without correspondingly large explicit breaking today.

However, this approach is disrupted by the inclusion of SUSY breaking.  For instance, if SUSY breaking is realized via a spurion superfield $\xi = M_P + \theta^2 F_\xi$, where $\abs{F_\xi}^2 \sim m_{3/2}^2 M_P^2$, then operators such as
\be
\Delta K = \abs{\xi^\dagger \xi} \left(\frac{k m_I^\dagger S^N}{M_P^{N+1}} + \hc\right),
\ee
will generate terms in the scalar potential
\be \label{eqn:suppressedSN}
V \supset \frac{k m_I^\dagger m_{3/2}^2 S^N}{M_P^{N-1}} + \hc
\ee
Even with the additional $m_{3/2}$ suppression, $N$ is constrained to be relatively large by strong CP considerations \cite{Dobrescu:1996jp}, reducing the efficacy of the $\abs{I}^2 S^N$ operator in giving mass to the axion during inflation.
For instance we find that, for low-scale SUSY breaking $m_{3/2} \sim \tev$ and $m_I \simeq 10^{-5} M_P$, the values of $N$ required preclude $m_{\rm eff, inf}^2 > H_I^2$ for inflationary parameters such as those considered above if $\vev{S} \sim f_a$ during inflation.

Additional parameter space would likely open if $\vev{S}$ were boosted during inflation, but such a model would require a mechanism for generating the displacement and gains would be limited relative to the model of \Sref{sec:expbreak}.  Indeed, even for the model above, it would be necessary to explain why the PQ fields exhibited $\vev{S} \sim \vev{\bar{S}} \sim f_a$, as opposed to, \eg, a minimum in which $\vev{S}$ was such that the effective inflaton mass vanished.

So, while the large value of the inflaton makes it seemingly an ideal candidate for enhancing explicit PQ breaking during inflation, it is in fact difficult to generate the desired operators without residual dangerous contributions to the axion mass.  This motivates considering whether the presence of another field, which still acquires a large vev during inflation but whose nature and symmetry properties are not as constrained as those of the inflaton, might be able to enhance explicit breaking during inflation.

\subsection{Additional $U(1)_{\rm PQ}$ Fields}

A challenge faced by models with a single PQ field is that the explicit breaking experienced by the axion pNGB during inflation, responsible for generating the large mass, is necessarily related to the explicit breaking today, which is constrained to be small by the strong CP problem.
However, if there are multiple fields that contribute to the breaking of $U(1)_{\rm PQ}$, the QCD axion 
will be a linear combination of the axial components of these fields.  As such, strong CP constraints on each field will depend on the amount of the axion contained within that field.
If the identity of the axion changes with time, it could ``feel'' more explicit breaking during inflation than today without running afoul of strong CP constraints.  

To get a sense of the issues one must consider in constructing a viable model, we consider a ``toy'' consisting of two PQ fields, $S$ and $\bar{S}$, with charges $q_S = 1, q_{\bar{S}} = -K$ under $U(1)_{\rm PQ}$, which is explicitly broken to $\mathbb{Z}_{K N}$.  As such, the lowest dimension Planck-suppressed local operators are
\be
\label{eq:twoPQinv}
V \supset \frac{k \bar{S} S^K}{M_P^{K-3}} + \frac{k' \bar{S}^{N}}{M_P^{N-4}} + \frac{k'' S^{K N}}{M_P^{K N - 4}} + \hc
\ee
In the limit where $U(1)_{\rm PQ}$ is not explicitly broken ($k' = k'' = 0$), the QCD axion and PQ breaking scale are
\begin{align}
\label{eq:aQCD}
a_{\rm QCD} & = \frac{1}{v_a} \sum_{i = S, \bar{S}} q_i v_i a_i, & v_a^2 & = \sum_{i = S, \bar{S}} q_i^2 v_i^2,
\end{align}
where $a_S$ denotes the axial component of $S$ and $\vev{S} \equiv v_S$, and similar for $a_{\bar{S}}, v_{\bar{S}}$.
The $\bar{S}^N$ and $S^{K N}$ operators both contribute to a mass for the axion, with the dominant contribution coming from the former.
$\bar{S} S^K$ is $U(1)_{\rm PQ}$ invariant and so, while it gives mass to the orthogonal axial field $a_\perp$, it only contributes to $m_{\rm eff}^2$ via heavily suppressed mass mixing.

The vevs of $S$ and $\bar{S}$ are set by the potential,
\begin{align}
\label{eq:SbarSvevpotential}
\begin{split}
V \supset ~ & \lambda_S \left(\abs{S}^2 - \frac{f_S^2}{2}\right)^2 + m_{\bar{S}}^2 \abs{\bar{S}}^2 + \frac{\lambda_{\bar{S}}}{4} \abs{\bar{S}}^4 \\
& - \frac{\delta_{\bar{S}}}{2} I^2 \abs{\bar{S}}^2.
\end{split}
\end{align}
Today, when $\vev{I} = 0$ is assumed, $v_{\bar{S},0} \neq 0$ is driven by the tadpole for $\bar{S}$ induced via the first term of \Eref{eq:twoPQinv} when $v_{S,0} \simeq \frac{f_S}{\sqrt{2}}$. For instance, neglecting the terms with powers of $\bar{S}$ greater than two,\footnote{\label{fn:neglectquartic} The quartic term can be neglected for $\lambda_{\bar{S}} \ll \frac{2 m_{\bar{S}}^6 M_P^{2K - 6}}{\abs{k}^2 f_S^{2K}}$.}
\be
\label{eq:Sbartoday}
\vev{\bar{S}}_0 \equiv v_{\bar{S}, 0} \simeq \frac{k v_{S,0}^K}{m_{\bar{S}}^2 M_P^{K - 3}}.
\ee
Thus, $v_{\bar{S},0}$ can be considerably smaller than $v_{S, 0}$ due to the $\left(\frac{v_{S,0}}{M_P}\right)^{K-3}$ suppression, supposing $m_{\bar{S}}^2 \slashed{\ll} v_{S, 0}^2$.  In this case, the present day axion resides predominantly in $S$ and $v_a \simeq v_{S,0}$, allowing $N$ significantly smaller than $K N$ consistent with strong CP constraints.

Meanwhile, during inflation, $v_{\bar{S}}$ is enhanced to $v_{\bar{S}, I} > v_{S,I}$ by the coupling to the inflaton in the second line of \Eref{eq:SbarSvevpotential}.
However, we assume $v_{S,I} \simeq v_{S,0} \simeq \frac{f_S}{\sqrt{2}}$ by taking $\lambda_S$ large enough that the cross coupling for $S$, $I^2 \abs{S}^2$, is negligible.\footnote{Similarly, we neglect terms such as $\abs{S}^2 \abs{\bar{S}}^2$.  For constant $\vev{S}$, this term can be considered a contribution to $m_{\bar{S}}^2$.}
Because $v_{\bar{S},I} > v_{S,I}$, $a_{\rm QCD}$ is dominantly composed of $a_{\bar{S}}$ during inflation and so receives a large explicit PQ-breaking mass from $\bar{S}^N$.
The mass of $a_\perp \simeq a_S$ from the $\bar{S} S^K$ term is similarly enhanced by large $v_{\bar{S},I}$.
Therefore, all fields are heavy during inflation, eliminating isocurvature constraints.

\begin{figure}
\includegraphics[width=\columnwidth]{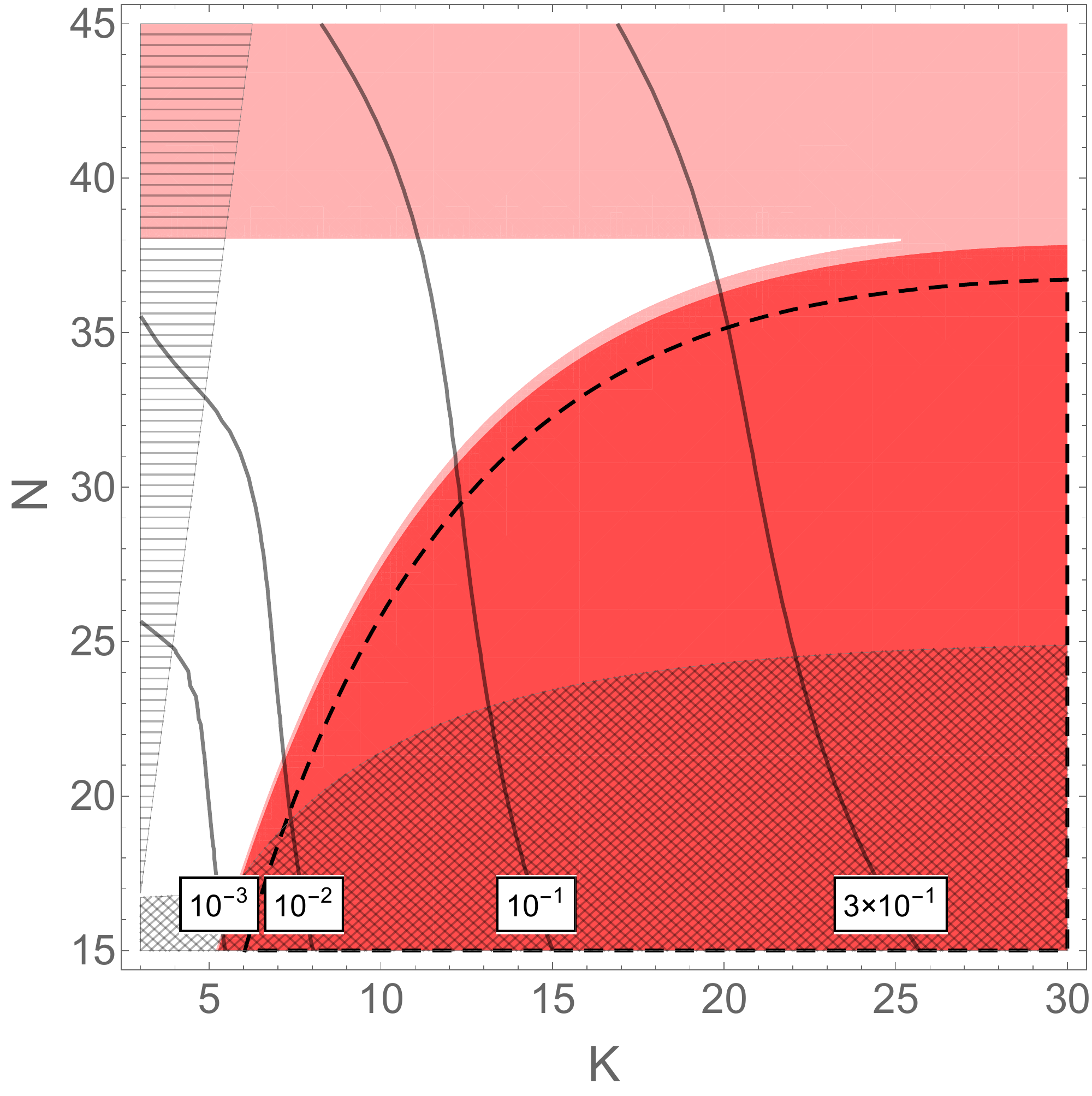}
\caption{\label{fig:SbarSNconstraints}
Bounds on multiple PQ field model for $\abs{k} = \abs{k'} = \abs{k''} = 1$, $m_{\bar{S}}^2= 10^{-11} M_P^2$.
Contours of the maximal $\frac{f_a}{M_P}$ allowed by strong CP constraints are solid black.
Regions where the contributions to the slow-roll parameters are greater than 1 (10) times current bounds are shaded light (dark) red.
The region where $v_{\bar{S},e} > 10^4 v_{\bar{S},0}$ so that symmetry restoration via parametric resonance may be a concern is crosshatched.
Within the dashed black contour denotes where our assumption that the $\lambda_{\bar{S}}$ term is negligible today breaks down.  In the horizontal-hatched region, $a_{\text{QCD}} \simeq a_{\bar{S}}$ always.
}
\end{figure}

The modification to the PQ potential required to enhance $v_{\bar{S}}$ during inflation still risks destabilizing the inflaton potential.  In \Fref{fig:SbarSNconstraints}, we show an example of the constraints on such a model with $\abs{k} = \abs{k'} = \abs{k''} = 1$ and $m_{\bar{S}}^2 = 10^{-11} M_P^2$.  For a given $K$ and $N$, we plot in black contours the maximal allowed value of $f_a \equiv \sqrt{2} v_a$ such that the contributions to the axion mass today due to explicit breaking do not disrupt the solution to the strong CP problem---\ie, both $N$ and $K N$ are sufficiently large that the $U(1)_{\rm PQ}$-breaking operators of \Eref{eq:twoPQinv} give small contributions to $m_{\rm eff}^2$.
As above, we implement strong CP constraints subject to the cosmological requirement that the minimum to which fields evolve after inflation is not significantly displaced from that favored by QCD (\ie, analogous to \Eref{eq:simplemeffconstraint}).

To simplify the analysis, we take $v_{\bar{S}, 0}$ to be given by \Eref{eq:Sbartoday}, which is of order $10^{-5}$ to $10^{-3}M_P$ in the allowed region.  As discussed below \Eref{eq:Sbartoday}, $v_{\bar{S},0}$ is not suppressed relative to $v_{S, 0}$ for small $K$, and thus $a_{\text{QCD}, 0} \simeq a_{\bar{S}}$---\ie, the composition of the axion is largely the same during inflation and today.  This results in dynamics essentially equivalent to those of the model of \Sref{sec:expbreak}.  As such, in this (horizontal-hatched) region there is no particular advantage to multiple PQ fields, though the model is viable.

Regions corresponding to (overly) large shifts to the slow-roll parameters in excess of 1 (10) times the current experimental constraints on the potential (\Eref{eqn:SNVpbound}) are shown in light (dark) red.\footnote{As above, $\delta_{\bar{S}}, \lambda_{\bar{S}}$ are conservatively fixed by the requirements that the potential is minimized at $\vev{\bar{S}} = v_{\bar{S}, I}$ and that the mass of the radial direction is sufficiently large $\simeq H_I$ that the field does indeed rapidly evolve to this minimum during inflation.}
Within the dashed curve, the tadpole approximation of \Eref{eq:Sbartoday} breaks down. 
Since this region is well excluded by fine-tuning considerations, we do not attempt to improve upon this approximation.

In the allowed region, $\frac{M_P}{5} \lsim f_{\bar{S}, I} = \sqrt{2} v_{\bar{S}, I} \lsim \frac{M_P}{2}$ is the minimum value required to give a sufficiently large mass $\gsim H_I$ to both $a_{\rm QCD}$ and $a_\perp$ (outside the allowed region it differs by less than an order of magnitude except in the darker red region).
The cross-hatched region denotes where the initial amplitude of the resulting $\bar{S}$ fluctuations is very large, $v_{\bar{S}, e} > 10^4 v_{\bar{S}, 0}$ (as before, subscript $e$ denotes the end of inflation).  Such large oscillations potentially produce large fluctuations in the phases of $S, \bar{S}$ that could result in additional contributions to the axion abundance, domain walls, or nonthermal symmetry restoration.
But this is not necessarily the case, and further investigation would be required to determine whether or not this region yields a consistent cosmology---see discussion in \Sref{sec:expbreak}.

\begin{figure}
\includegraphics[width=\columnwidth]{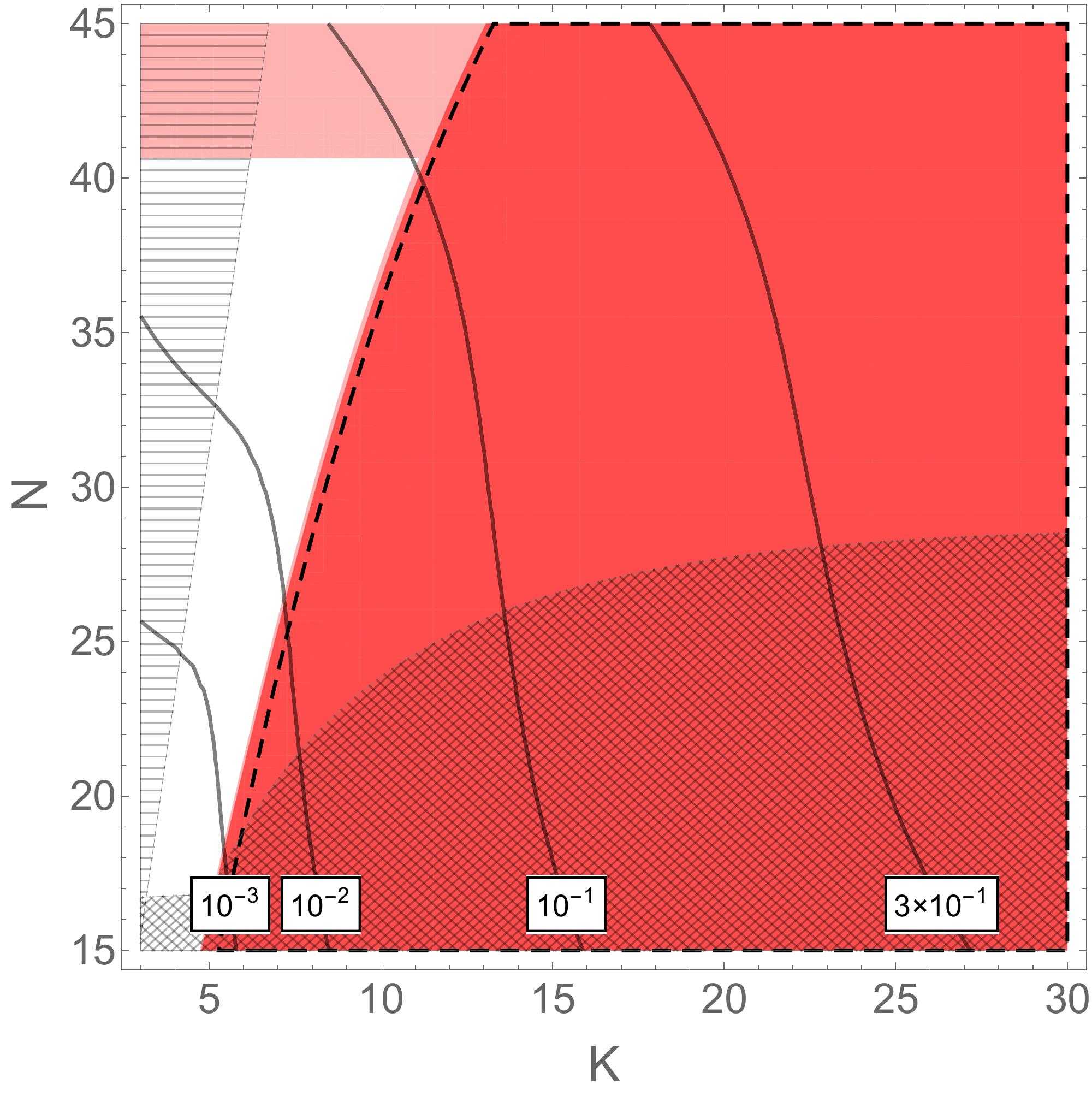}
\caption{\label{fig:SbarSNconstraints2}
Bounds on multiple PQ field model for $m_{\bar{S}}^2 = 10^{-12} M_P^2$.  All curves are the same as in \Fref{fig:SbarSNconstraints}.
}
\end{figure}

In \Fref{fig:SbarSNconstraints}, we have tuned $m_{\bar{S}}^2$ to be just less than $H_I^2$.
That this allows for near maximal reach in $f_a$ can be understood as follows.
The non-QCD contribution to the axion mass goes as $m_{\rm eff}^2 \propto \frac{v_{\bar{S},0}^N}{v_{S,0}^2} \propto \frac{f_S^{KN - 2}}{m_{\bar{S}}^{2 N}}$.
So, for smaller $m_{\bar{S}}^2$ and fixed $f_a$, $K$ and/or $N$ must be increased slightly to ensure that the contribution to $m_{\rm eff}^2$ is sufficiently small today.
However, small increases in $K$ exponentially increase the necessary $v_{\bar{S}, I}$ required to give a large enough mass to $a_\perp$ during inflation, since its mass is proportional to only a single power of $\bar{S}$.
Thus, the bottom right red region moves significantly to the left while the black contours move slightly to the right---when taken together, this can drastically reduce the values of $f_a$ achievable.
Along similar lines, slightly larger $f_a$ may be accommodated by tuning $m_{\bar{S}}^2$ marginally closer to $H_I^2$.
But gains are limited because, as $m_{\bar{S}}^2$ approaches (or exceeds) $H_I^2$, larger $\delta_{\bar{S}}$ is also eventually required to overcome the positive mass-squared parameter, such that slow-roll constraints rapidly exclude the whole parameter space.
Thus, \Fref{fig:SbarSNconstraints} represents close to the maximal (rather than typical) reach for a model of this type.
For comparison, in \Fref{fig:SbarSNconstraints2} we show the analog of \Fref{fig:SbarSNconstraints} for $m_{\bar{S}}^2 = 10^{-12} M_P^2$.  As $m_{\bar{S}}^2$ decreases, the region consistent with slow-roll constraints moves to overlap with the horizontal-hatched region where $a_{\rm QCD} \simeq a_{\bar{S}}$ always, with (horizontal) slow-roll constraints excluding $N \gsim 40$ in agreement with \Sref{sec:expbreak}.

Further parameter space could also be reached if $v_{S, I} > v_{S, 0}$.  Such a setup requires a mechanism for boosting $\vev{S}$ as well.
Constraints on $N$ would vanish entirely for $v_{\bar{S}, 0} = 0$, which could perhaps be achieved for different $q_{S, \bar{S}}$, but for vanishing vev it is difficult to achieve efficient $\bar{S}$ decay.
For instance, if $\bar{S}$ couples to PQ quarks via $\bar{S} Q \bar{Q}$, couplings such as $S^M Q \bar{Q}$ are forbidden, precluding the quarks from acquiring mass.

As for the models explored in the previous subsection, this setup with multiple PQ fields and changing vevs can open parameter space at larger $f_a$, potentially even allowing models with $f_a \sim M_P$.
However, for these largest values, remaining consistent with isocurvature constraints while avoiding excessive disruption of inflaton dynamics is still difficult and requires some conspiracy between $K, N, m_{\bar{S}}^2$ and inflationary parameters.
Similar model building may even allow $f_a \gsim M_P$.  But, in this regime, the analysis presented here may well be insufficient as higher-order operators, no longer necessarily suppressed, could play a role.  Moreover, $f_a \gsim 10^{-1} M_P$ is excluded by bounds from black hole superradiance \cite{Arvanitaki:2010sy}.

\section{Additional Constraints on Radial Field Displacement}
\label{sec:additionalconstraints}

There are various other constraints that we have not so far considered when the radial field is significantly displaced during inflation from its present minimum.  Mostly these are model dependent and can be avoided under appropriate circumstances that we detail.

\subsection{Perturbations in the Radial Field}
\label{sec:radialperturbations}

A light PQ field that was displaced due to Hubble friction would obtain primordial fluctuations $\delta \sigma = \frac{H_I}{2 \pi}$, which are orthogonal to the curvature perturbations seeded by the inflaton.
The decay products of $\sigma$ will inherit these fluctuations, potentially providing an additional source of isocurvature.

This may be erased if local thermal equilibrium is achieved \cite{Weinberg:2004kf}.
Alternatively, if the radial mode comes to dominate the energy density and its decay products reheat the universe, it will effectively act as a ``curvaton'' \cite{Lyth:2001nq}---the observed perturbations result from $\sigma$ fluctuations as opposed to those of a separate inflaton, so are not observed as isocurvature modes.
If $\sigma$ decays when its energy density is subdominant, it can still seed the observed curvature perturbations in a curvaton-like fashion if inflationary curvature perturbations are subdominant or absent.  Although, in this case, the perturbations arising from $\sigma$ must be larger to compensate for its subdominance, which frequently results in sizable non-Gaussianities, see \eg~\cite{Beltran:2008aa}.
In the event that $\sigma$ does indeed decay while subdominant and is not a curvaton, exact constraints depend on the epoch---for a detailed analysis in the context of moduli, see \cite{Iliesiu:2013rqa}.

\subsection{Scalar Trapping}
\label{sec:scalartrapping}

If the radial field is responsible for giving the PQ quarks (and squarks, if present) mass, symmetry restoration may occur as a result of scalar trapping as described in \cite{Moroi:2013tea}.  This trapping proceeds as follows.  When $\abs{S}$ becomes small during its oscillations, the PQ (s)quarks become light enough to be produced via thermal effects or parametric resonance.  These (s)quarks then backreact on $\abs{S}$, leading to it becoming trapped at the origin and thus effectively restoring the PQ symmetry.  For this to occur, the initial oscillation amplitude must be large enough that $\abs{S}$ passes near the origin, which only occurs for $\abs{S}_i \gsim 10 \frac{f_a}{\sqrt{2}}$.  Scalar trapping may be avoided if there are no PQ squarks, there are additional sources of mass for the PQ (s)quarks (\eg, in multifield PQ models), or the radial field is sufficiently heavy that it does not spend a significant time in the critical regime during each oscillation.

\subsection{Radial Field Energy Density}
\label{sec:radialenergydensity}

As the $\sigma$ oscillations decay, the corresponding energy density is transferred into its decay products, leading to constraints that depend in detail upon when the decay occurs and to what final state(s) \cite{Kawasaki:2007mk}.
If it decays to axions, they act like dark radiation, which is constrained by measurements of the CMB and the success of BBN.
Alternatively, the radial field can decay predominantly to colored states (including gluons or PQ quarks), which would avoid dark radiation constraints.
The decays of the radial field must not disrupt BBN, so it must either decay before the start of BBN or remain a (substantially) subdominant component of the energy density.
Both conditions favor earlier decay of the radial field---before BBN or before its proportional energy density has increased significantly---which generally corresponds to heavier PQ fields.
Such a heavy field is possible in a generic model, but could pose a problem in supersymmetric models where there is necessarily a light saxion field with mass of order the SUSY-breaking scale due to the complexification of the $ U(1)_{\rm PQ}$ symmetry \cite{Kugo:1983ma}.

Decays before BBN can still be constrained by the observed DM relic density if $\sigma$ decays to DM after thermal freeze-out.  Again, this constraint may be more difficult to avoid in supersymmetric modes with R-parity, \eg, for a saxion decaying to superpartners whose decays yield stable lightest supersymmetric particles.  The extent to which this is a concern depends on whether supergravity effects yield $\text{Br}(\sigma \rightarrow \text{ gauge bosons}) \simeq \text{Br}(\sigma \rightarrow \text{ gauginos})$, as suggested in \cite{Endo:2006ix}, in which case a sizable branching ratio to gluons (desired to avoid dark radiation constraints) would be accompanied by a sizable ratio to gluinos.  However, other analyses suggest that decays to gauginos suffer from additional chiral suppression (\ie, by the mass of the gaugino) \cite{Baer:2010gr}, in which case decays to superpartners would be subdominant.

\section{Can the Radial Field Evolve Adiabatically?}
\label{sec:adiabatic}

Constraints arising from $\sigma$ oscillations (both thermal and nonthermal) as discussed in the previous sections would be evaded if $\vev{S}$ evolved gradually to $f = f_a$.  This can only occur if the PQ potential does not change violently at the end of inflation, which does not occur for generic couplings between the PQ and inflationary sectors.  Typically, the end of inflation corresponds to a radical change in dynamics in the inflationary sector---in slow-roll models, the inflaton leaves the slow-roll regime and begins to coherently oscillate around the minimum of its potential---and as such a similarly drastic change to the PQ potential is to be expected.  In all likelihood, the mechanism responsible for maintaining $\vev{S} = \frac{f_I}{\sqrt{2}}$ rapidly disappears and the radial component of the PQ field begins oscillating around the zero temperature minimum $\vev{S} = \frac{f_a}{\sqrt{2}}$.

However, if the size of the operator coupling the PQ and inflationary sectors decreased gradually, $\sigma$ and $\vev{S}$ could conceivably evolve adiabatically to a lower value.  The obvious candidate for such a solution is for $S$ to couple to the full energy density of the inflaton $\rho_I$, which is nonoscillatory but rather decreases gradually up until inflaton decay.  An analogous approach has been considered previously to alleviate constraints on moduli energy density through assuming moduli couple directly to $H^2 \propto \rho_{\rm total}$ (which is equal to the inflaton energy density prior to reheating) \cite{Linde:1996cx}.  Couplings proportional to $H$ have also been invoked to reduce the amplitude of saxion oscillations \cite{Kawasaki:2011aa} and in the context of models to suppress axion isocurvature via a sufficiently high $f_I$ \cite{Chun:2004gx,Chun:2014xva}.

Unfortunately, there are two obstacles to invoking such a solution.  First, even supposing a coupling such as $c H^2 \abs{S}^2$ does dominate, coherent oscillations are generally diminished but not completely avoided.
In a SUSY model, for instance, the minimum today is determined by the soft masses, so is likely displaced $\gsim f_a$ from the minimum preferred by ${\cal O}(H^2)$ masses (see, \eg, \cite{Kawasaki:2007mk}).
Moreover, in a general model, adiabatic tracking reduces the initial amplitude of oscillations, but oscillations still commence eventually, with the extent of the reduction depending on the magnitude of $c$ and the other terms in the $S$ potential \cite{Linde:1996cx,Nakayama:2011wqa,Harigaya:2015hha}.
Suppressing the initial amplitude by even an order of magnitude requires $c \gsim {\cal O}(10)$ \cite{Linde:1996cx,Nakayama:2011wqa}---\eg, for $c = (4 \pi)^2$ and a quartic PQ potential, the initial oscillation amplitude is only reduced by a factor of $\sim 10$ \cite{Harigaya:2015hha}.
As alluded to in previous sections, too large $c$ risks interfering with inflationary dynamics, particularly for models requiring (super-)Planckian $f_I$.

Second, implementing such a coupling to $\rho_I$  appears difficult from a model-building standpoint, as it requires tuning between the couplings of the PQ field and the inflaton kinetic and potential terms.  We elaborate on this issue below, but stress again that even if such a model can be successfully constructed it only realistically reduces the oscillation amplitude by a factor of ${\cal O}(10)$.  So, at the end of the day, the constraints of the previous sections are likely still significant.

\subsection*{Implementing Adiabatic Relaxation of the PQ Field}

As mentioned, at the most basic level, coupling to a nonoscillatory or smoothly varying quantity requires a tuning between the couplings of the PQ field to the inflaton kinetic and potential terms.  One might hope to invoke a symmetry to enforce the required coincidence; a frequently considered candidate is supersymmetry.  For instance, a SUSY inflationary model containing a K\"ahler potential coupling
\be
\label{eq:Kahlerc}
K \supset - c \frac{\abs{I}^2 \abs{S}^2}{M_P^2},
\ee
where $I$ is an inflaton field whose potential is dominated by its $F$-term, $V_I \simeq \abs{F_I}^2$, exhibits a coupling between $\abs{S}$ and the energy density of $I$ as $\vevnoabs{\int d^4 \theta \abs{I}^2} = \rho_I$ \cite{Dine:1995uk,Dine:1995kz},
\be
{\cal L} \supset - \frac{c}{M_P^2} \left(\abs{\partial_\mu I}^2 + V_I\right) \abs{S}^2 = - 3 c H^2 \abs{S}^2,
\ee
where the last equality holds when $\rho_I$ is the dominant component of the energy density of the universe.
Supposing other operators were subdominant, the evolution of $\vev{S}$ would be determined by the gradually red-shifting $\rho_I$.  Moreover, for $c \gg 1$, the highly curved nature of the potential in the vicinity of the vev would cause $S$ to evolve rapidly to the minimum, such that the field approximately tracked the adiabatically evolving vev without large oscillations.

However, it is not obviously sufficient to consider only these terms---for instance, as observed in \cite{Dine:1995kz}, a number of comparably sized operators will generically arise from various sources.  In particular, SUGRA corrections will generate additional couplings that disrupt the relationship between the coupling of $S$ to inflaton kinetic and potential terms.  
Notably, expanding the exponential in \Eref{eq:SUGRAV} will lead to additional couplings of $S$ to $I$ in the scalar potential without corresponding couplings to the kinetic terms for $I$.

One might worry that these terms would spoil the success of this solution in suppressing large coherent oscillations.  Indeed, when $I$ starts to oscillate, the additional terms, which necessarily play a role in determining $\vev{S}$, would rapidly change or disappear.  As such, the vev would also rapidly change and so $S$ too would be expected to start oscillating.
But, for the large values of $c \gg 1$ required for this adiabatic tracking mechanism to work, the coupling proportional to $H^2$ may dominate, in which case the change to $\vev{S}$ could be small.  Furthermore, the coherent oscillations would occur around the new nearby minimum as opposed to the minimum today and would be rapidly damped due to the large effective mass $\sim \sqrt{c} H$.
So, small perturbations introduced by couplings to the oscillating inflaton field $I$ as opposed to $H^2$ do not necessarily disrupt the tracking.
However, the post-inflationary dynamics of both the inflaton and PQ fields are potentially complicated in this scenario, and depend on their coupled equations of motion.  The exact behavior will depend on model-specific details and an analysis of the viability of this solution in well-motivated examples is an interesting question, albeit beyond the scope of this work.

A more serious concern is that the absence of tuning permitted by SUSY in this case---\ie, that a single term can generate the desired coupling to $\rho_I \propto H^2$---relies on the inflaton potential being dominated by its $F$-term, $F_I$.  
As is well known, SUSY inflationary models of this type exhibit a severe $\eta$ problem, namely that $\eta \simeq {\cal O}(1)$ if various contributions are not tuned against one another (for a review see, \eg, \cite{Lyth:1998xn,Baumann:2009ds})---notably, sizable contributions to $\eta$ arise from the exponential in \Eref{eq:SUGRAV}.  The $\eta$ problem is potentially exacerbated in this case, especially for $c \gg 1$ and $\abs{S} \sim M_P$.
So, while tuning might not be required to get the desired coupling between the inflaton and PQ field, it may still be necessary to yield a viable model of inflation.

On the other hand, models do exist in which the $\eta$ problem is solved without tuning by an additional symmetry for the inflaton, such as the model considered in \Sref{sec:ISNmodel}.  Then, though, the inflaton potential is not dominated by $F_I$ and so additional tuning would be required to achieve a dominant coupling to $H^2$.  For instance, in the model discussed previously with
\be
K \supset \abs{X}^2 + \frac{1}{2} \left(I + I^\dagger\right)^2, \quad W = m_I X I,
\ee
the shift symmetry $I \rightarrow I + i c$ prevents the imaginary scalar component of $I$, $\phi_I$, from appearing nonderivatively from the K\"ahler potential.  As such, $\phi_I$ can take on large field values without an associated $\eta$ problem, allowing it to act as the inflaton while the heavy real component of $I$ and the $X$ scalar are stabilized at the origin.
In this model, though, inflation is driven by the $F$-term for $X$ rather than that for $I$,
\be
V_F \simeq \abs{F_X}^2 = \frac{m_I^2}{2} \phi_I^2,
\ee
where we have taken $X = \text{Re}(I) = 0$.  So, equal coupling to the kinetic and potential terms for $\phi_I$ requires multiple additional terms in the K\"ahler potential
\be
K \supset \abs{S}^2 - \frac{c}{2 M_P^2} (I + I^\dagger)^2 \abs{S}^2 - \frac{d}{M_P^2} \abs{X}^2 \abs{S}^2
\ee
and a tuning $c \simeq d - 1 \gg 1$.  Moreover, higher order terms such as $\frac{\abs{S}^4 \abs{F_X}^2}{M_P^4}$ are still generated, which may affect the dynamics of the fields after inflation.

Furthermore, a similar ``$\eta$'' problem exists for the PQ field, making it difficult to stabilize $\vev{S}$ at a large field value.  Specifically, the exponential barrier tends to drive $\vev{S} \rightarrow 0$ or $\vev{S} \rightarrow \infty$ during inflation.
Unlike for a generic modulus, a shift symmetry solution cannot be invoked for $S$ to prevent it from appearing explicitly in $K$ as such a symmetry is incompatible with $U(1)_{\rm PQ}$.
The tunings required to avoid the inflaton $\eta$ problem are not the same as those required to generate the desired potential for $S$, such that the solutions to these problems are not necessarily related.

Both problems are more severe for large-field models as the field values $I_\ast \gsim M_P$ during inflation mean higher-order operators are effectively unsuppressed.  As these are exactly the models that yield observable scalar-to-tensor ratios and for which axion isocurvature is a major concern, it is unclear that an adiabatic solution can be readily implemented to suppress isocurvature without large coherent oscillations.  At the very least, both a precise tuning of various terms and large couplings appear necessary to simultaneously achieve inflation, the desired inflationary minimum, and adiabatic tracking behavior, even with supersymmetry.


\section{Conclusions}
\label{sec:conclusion}

In models where the PQ symmetry breaks before the end of inflation, a high scale of inflation  na\"ively induces too large isocurvature perturbations in the CMB.  However, if the PQ sector is modified from its zero temperature form during inflation, this need not be the case.  

One possibility is that, during inflation, the PQ field is still evolving towards its minimum from a large initial value.  This occurs, for example, in a model wherein the radial component of the PQ field acts as the inflaton.
In this scenario, however, post-inflationary radial oscillations can potentially induce the restoration of the PQ symmetry, with disastrous cosmological consequences. Indeed, the ``PQ sector inflation" approach only seems viable if there is a late-time release of entropy. Even then, this is only the case for a narrow window $f_a \sim 10^{-3} M_P$, and more detailed studies would be required to ensure symmetry restoration would not occur.
Moreover, as constraints on isocurvature tighten, this class of solution will become increasingly untenable.

Another approach is to induce modifications of the PQ potential by explicitly breaking the PQ symmetry and coupling the PQ sector to the inflaton.  However, this has its own set of challenges. The inflaton potential is necessarily delicate---it must satisfy the slow-roll conditions.  Thus, maintaining a sufficiently flat potential in the presence of large couplings between the PQ sector and the inflation sector can necessitate aesthetically unpleasing fine-tunings.
And the axion potential must be dominated by QCD---additional contributions run the risk of spoiling the elegant  solution to the strong CP problem.  With judicious choice of potential, it is indeed possible to avoid these concerns, and GUT-scale or even Planck-scale $f_{a}$ is allowed.  These models typically rely on a discrete symmetry,  with a delicate explicit breaking which is amplified during the inflationary period. Furthermore, this solution does not allow a small axion abundance via an anthropically chosen initial misalignment angle $\theta_i$, so either a coincidence in the phase of the operator, or a late-time dilution (perhaps due to a modulus) is required.

Given the fragility of both the inflaton and the axion, it might be productive to introduce yet another field that amplifies the PQ breaking.  This could insulate the inflaton against the fine-tuning effects discussed above.  One possibility might be a modulus that obtains a large vev during inflation.   However, potential reintroduction of the strong-CP problem by, \eg, generation of a tadpole, is still a concern.  The viability of this solution is an interesting direction for further work.

Even if isocurvature is suppressed by one of the above mechanisms or similar, precautions must be taken to ensure a consistent cosmology.
This is particularly true if the transition to the field configuration today is violent, involving large coherent field oscillations or significant energy density stored in late-decaying fields.
While these challenges are well known from the physics of moduli and saxions and can be evaded by appropriate model building, they must be taken into account and may well constitute the dominant constraints, especially as coherent oscillations can perhaps be reduced but generally not eliminated.
Additional fields (multiple PQ fields, separate inflatons, moduli) offer more alternatives for suppressing isocurvature, but their nontrivial post-inflationary dynamics will also yield additional constraints.

If primordial tensor perturbations are indeed observed in the CMB, models with large $f_{a}$ will require some additional physics coupled to the PQ sector. 
If residual isocurvature is also observed, it may be the case that isocurvature is simply suppressed by $f_{I} \gg f_{a}$.  Alternatively, if no isocurvature is visible, it may be that a large mass for the axion is generated during inflation.
In either case, future precision probes of the CMB and axion DM stand to tell us much about high-scale axions, the inflationary sector, and, perhaps, an interplay between the two.

\acknowledgments{We would like to thank Yue Zhao, Patrick Fox, Kenji Kadota and Kiel Howe for valuable discussions.  JK is supported by the DoE under contract number DE-SC0007859 and Fermilab, operated by Fermi Research Alliance, LLC under contract number DE-AC02-07CH11359 with the United States Department of Energy.  The work of AP and NO is supported by the U.S. Department of Energy under grant DE-SC0007859.}

\bibliography{axionrefs}

\end{document}